%% file: main.tex
\documentclass[aps,prd,reprint,groupedaddress,nofootinbib,floatfix]{revtex4-2}

\input{structure.tex}

\begin{document}

\title{Leptonic signatures of color-sextet scalars}

\author{Linda M. Carpenter}
\email{lmc@physics.osu.edu}
\author{Taylor Murphy}
\email{murphy.1573@osu.edu}
\author{Katherine Schwind}
\email{schwind.44@osu.edu}
\affiliation{Department of Physics, The Ohio State University\\ 191 W. Woodruff Ave., Columbus, OH 43212, U.S.A.}

\date{\today}

\input{TeX/abstract}

\maketitle

\input{TeX/1_Intro}
\input{TeX/2_Model}
\input{TeX/3_LHC}
\input{TeX/4_CMS-EXO-17-009}
\input{TeX/5_Projections}
\input{TeX/6_Conclusion}


\appendix

\input{TeX/A_MG5}
\input{TeX/B_Unitarity}

\acknowledgments

This work was supported by the Department of Physics of The Ohio State University.


\bibliographystyle{apsrev4-2}
\bibliography{Bibliography/bibliography.bib}

\end{document}

%% file: structure.tex
\usepackage[svgnames]{xcolor} 

\usepackage[printonlyused,withpage]{acronym} 
\usepackage{amsfonts}
\usepackage{anyfontsize}
\usepackage{bbm}
\usepackage{booktabs} 
\usepackage{braket}
\usepackage{cancel}
\usepackage{changepage}
\usepackage{colortbl}
\usepackage{enumitem}
\usepackage[mathscr]{euscript}
\usepackage{float}
\usepackage[T1]{fontenc} 
\usepackage{gensymb}
\usepackage[hidelinks]{hyperref} 
\usepackage{listings} 
\usepackage{makecell}
\usepackage{mathtools}
\usepackage[framemethod=tikz]{mdframed} 
\usepackage{microtype} 
\usepackage{multirow}
\usepackage{physics}
\usepackage{relsize}
\usepackage{rotating} 
\usepackage{sansmath}
\usepackage{simplewick}
\usepackage{slashed}
\usepackage{stmaryrd}
\usepackage{subfig}
\usepackage{tcolorbox}
\usepackage{theorem}
\usepackage{tikz}
\usepackage{tikz-feynman,contour}
\usepackage[normalem]{ulem}
\usepackage{xspace} 
\usepackage{yfonts}
\usepackage[vcentermath]{youngtab}

\usepackage{newtxtext,newtxmath}

\usepackage{siunitx}

\tcbuselibrary{theorems}
\usetikzlibrary{decorations.markings}
\usetikzlibrary{shapes.geometric}

\def\d{{\mathrm{d}}} 
\def\e{{\text{e}}}
\def\ii{{\text{i}}}

\makeatletter
\newcommand*{\transpose}{%
  {\mathpalette\@transpose{}}%
}
\newcommand*{\@transpose}[2]{%
  \raisebox{\depth}{$\m@th#1\intercal$}%
}
\makeatother

\newtcbox{\sln}{colback=Gainsboro,
colframe=Gainsboro}

\newcommand{\tv}[1]{\overset{{}_{\,\scalebox{0.55}{$\shortrightarrow$}}}{#1}}

\newcommand{\bt}[1]{{\sansmath{\boldsymbol{#1}}}}

\newcommand{\overbar}[1]{\mkern 2mu\overline{\mkern-4mu#1\mkern-4mu}\mkern 2mu}

\tikzset{snake it/.style={decorate, decoration={snake,amplitude=10mm}}}

\tikzset{/pgf/decoration/.cd,
    number of sines/.initial=10,
    angle step/.initial=20,
}

\newdimen\tmpdimen

\pgfdeclaredecoration{full sines}{initial}
{
    \state{initial}[
        width=+0pt,
        next state=move,
        persistent precomputation={
            \pgfmathparse{\pgfkeysvalueof{/pgf/decoration/angle step}}%
            \let\anglestep=\pgfmathresult%
            \let\currentangle=\pgfmathresult%
            \pgfmathsetlengthmacro{\pointsperanglestep}%
                {(\pgfdecoratedremainingdistance/\pgfkeysvalueof{/pgf/decoration/number of sines})/360*\anglestep}%
        }] {}
    \state{move}[width=+\pointsperanglestep, next state=draw]{
        \pgfpathmoveto{\pgfpointorigin}
    }
    \state{draw}[width=+\pointsperanglestep, switch if less than=1.25*\pointsperanglestep to final, 
        persistent postcomputation={
        \pgfmathparse{mod(\currentangle+\anglestep, 360)}%
        \let\currentangle=\pgfmathresult%
    }]{%
        \pgfmathsin{+\currentangle}%
        \tmpdimen=\pgfdecorationsegmentamplitude%
        \tmpdimen=\pgfmathresult\tmpdimen%
        \divide\tmpdimen by2\relax%
        \pgfpathlineto{\pgfqpoint{0pt}{\tmpdimen}}%
    }
    \state{final}{
        \ifdim\pgfdecoratedremainingdistance>0pt\relax
            \pgfpathlineto{\pgfpointdecoratedpathlast}
        \fi
   }
}

\pgfdeclaredecoration{completely sines}{initial}
{
    \state{initial}[
        width=+0pt,
        next state=upsine,
        persistent precomputation={\pgfmathsetmacro\matchinglength{
            \pgfdecoratedinputsegmentlength / int(\pgfdecoratedinputsegmentlength/\pgfdecorationsegmentlength)}
            \setlength{\pgfdecorationsegmentlength}{\matchinglength pt}
        }] {}
    \state{upsine}[width=\pgfdecorationsegmentlength,next state=downsine]{
        \pgfpathsine{\pgfpoint{0.25\pgfdecorationsegmentlength}{0.5\pgfdecorationsegmentamplitude}}
        \pgfpathcosine{\pgfpoint{0.25\pgfdecorationsegmentlength}{-0.5\pgfdecorationsegmentamplitude}}
    }
    \state{downsine}[width=\pgfdecorationsegmentlength,next state=upsine]{
        \pgfpathsine{\pgfpoint{0.25\pgfdecorationsegmentlength}{-0.5\pgfdecorationsegmentamplitude}}
        \pgfpathcosine{\pgfpoint{0.25\pgfdecorationsegmentlength}{0.5\pgfdecorationsegmentamplitude}}
}
    \state{final}{}
}

\tikzfeynmanset{ with arrow/.style = {
   decoration={
     markings,
     mark=at position 0.5
          with {\arrow[xshift=1.9mm]{Latex[width=2.75mm,length=3mm]}}
     },
   postaction=decorate}
}

\tikzfeynmanset{ with glarrow/.style = {
   decoration={
     markings,
     mark=at position 0.5
          with {\arrow[white,xshift=3mm]{Latex[width=4.125mm,length=4.5mm]}}
     },
   postaction=decorate}
}

\tikzfeynmanset{ with outarrow/.style = {
   decoration={
     markings,
     mark=at position 0.5
          with {\arrow[xshift=3.35mm]{Latex[width=4.58333mm,length=5mm]}}
     },
   postaction=decorate}
}

\tikzfeynmanset{ bigphoton/.style={
    /tikz/draw=none,
    /tikz/decoration={name=none},
    /tikz/postaction={
      /tikz/draw,
      /tikz/decoration={
        completely sines,
        segment length=3mm,
        amplitude=2.5mm,
      },
      /tikz/decorate=true,
    },
  },
}

\tikzfeynmanset{ roundbigphoton/.style={
    /tikz/draw=none,
    /tikz/decoration={name=none},
    /tikz/postaction={
      /tikz/draw,
      /tikz/decoration={
        full sines,
        segment length=3mm,
        amplitude=1.25mm,
      },
      /tikz/decorate=true,
    },
  },
}

\tikzset{ mega thick/.style= {line width = 3.4pt}
}

\makeatletter
\renewcommand{\fnum@figure}{\textsc{\figurename~\thefigure}} 
\makeatother

\lstnewenvironment{pseudoc}{\lstset{frame=lines,language=C,mathescape=true}}{}
\lstnewenvironment{logs}{\lstset{frame=lines,basicstyle=\footnotesize\ttfamily,numbers=none}}{}
\lstnewenvironment{cc}{\lstset{frame=lines,language=C}}{}
\lstnewenvironment{c64}{\lstset{backgroundcolor=\color{c64},basewidth=0.65em,basicstyle=\commodoreface\color{c64light},numbers=none,framerule=10pt,rulecolor=\color{c64light},frame=tb,framexbottommargin=30pt}}{}
\lstnewenvironment{html}{\lstset{frame=lines,language=html,numbers=none}}{}
\lstnewenvironment{pseudo}{\lstset{frame=lines,mathescape=true,morekeywords={learn_string_domain, save_model}}}{}
\lstnewenvironment{pseudoctiny}{\lstset{language=C,mathescape=true,basicstyle=\tiny\sffamily}}{}
\lstnewenvironment{cctiny}{\lstset{language=C,basicstyle=\tiny\sffamily}}{}
\lstnewenvironment{pseudotiny}{\lstset{mathescape=true,basicstyle=\tiny\sffamily}}{}

%% file: TeX/abstract.tex
\begin{abstract}

Color-sextet scalars could have an array of possible couplings to the Standard Model beyond their well known renormalizable couplings to quark pairs. The next-largest couplings these scalars might enjoy have mass dimension six, and some include electroweak fields and can therefore produce highly distinctive signals at hadron colliders. Here we study the single production of color-sextet scalars in association with a hard lepton, which results from a dimension-six coupling to leptons in addition to quarks and gluons. We identify parameter space in which these scalars decay promptly and propose a search for such particles in final states with two jets and an opposite-charge lepton pair, one member of which has high momentum. We then impose our selection criteria on the proposed signal and its leading backgrounds to project the discovery potential and exclusion limits during the next runs of the LHC. We compare our search with existing searches for leptoquarks, and we find we achieve much higher sensitivity with our tailored approach due to the unique kinematics of leptonic sextet events.

\end{abstract}

%% file: TeX/1_Intro.tex
\section{Introduction}
\label{s1}

Among the many goals of the Large Hadron Collider (LHC) program is to leverage the increasingly large dataset --- $139\,\text{fb}^{-1}$ at the time of writing, and up to $3\,\text{ab}^{-1}$ after the high-luminosity (HL) upgrades \cite{https://doi.org/10.5170/cern-2015-005.1} --- to search for physics beyond the Standard Model (bSM). The utility of the LHC as a probe of bSM physics rests not only on the ATLAS and CMS collaborations, which so far have guided the collider through two runs with excellent performance and have designed searches for a variety of popular new-physics scenarios, but also on the wider community, which is constantly conceiving new models ripe for exploration at LHC and has developed tools to reinterpret existing analyses within new model frameworks \cite{Conte_2013,Conte_2014,Conte_2018}. The true potential of the LHC will only be realized through a combination of these parallel efforts, which are far from complete. In particular, there exists a formidable amount of bSM model parameter space yet to be investigated. Models featuring new states in exotic representations of the SM gauge group(s), and in particular the color gauge group $\mathrm{SU}(3)_{\text{c}}$, occupy some of this unexplored space. These states may have measurable effects on the electroweak (including Higgs) sector \cite{Boughezal:2010ry,Carpenter:2022oyg, Gisbert:2022lao, Manohar:2006ga} and produce an array of interesting collider signatures \cite{Chen:2014haa,Carpenter:2015gua,Carpenter:2011yj, Carpenter:2021gpl}.

Color sextets, which transform in the six-dimensional irreducible representation ($\boldsymbol{6}$) of $\mathrm{SU}(3)_{\text{c}}$, were first studied a few decades ago \cite{Chivukula_91,Celikel_98} and have garnered some renewed interest in the twenty-first century. The more recent works have focused on the so-called \emph{sextet diquarks} \cite{Han:2009ya}, color-sextet scalars enjoying renormalizable couplings to like-sign pairs of right-chiral (``handed'') SM quarks schematically of the form
\begin{align}\label{diquark}
    \mathcal{L} \sim \varphi^{\dagger} q_{\text{R}} q'_{\text{R}} + \text{H.c.}
\end{align}
with $q,q' \in \{u,d\}$ depending on the weak hypercharge of the sextet diquark $\varphi$. Some of the LHC phenomenology of these particles has been studied both in a standalone context \cite{PhysRevD.79.054002,Han:2010rf} and embedded as messengers between the Standard Model and a dark sector \cite{Carpenter:2022lhj}. However, so far only a small corner of the full space of couplings between sextets and the Standard Model has been probed. 

To wit, there exist a large number of heretofore unexplored effective operators that might induce couplings between color-sextet fields and the Standard Model and generate distinctive signals well suited for LHC searches. The operators describing these interactions have recently been systematically cataloged at mass dimensions five and six \cite{Carpenter:2021rkl}. The operator catalog is useful for revealing stand-out collider event topologies that open the way for new LHC searches. In particular, some of these operators are interesting because they include electroweak fields, most notably charged leptons, which are easy to track in the ATLAS and CMS detectors. In this work we explore a particular dimension-six interaction that couples a color-sextet scalar to a SM quark, a gluon, and a charged lepton. This Lorentz- and gauge-invariant operator permits single sextet scalar production at LHC in association with a hard lepton. In the absence of any other operators, this restricts the scalar to decay to (at least) a three-body final state. This chain of events generates multiple jets in addition to an opposite-sign lepton pair ($jj\,\ell^+\ell^-$), one lepton of which has particularly high momentum. We compare an array of lepton and jet observable distributions for our signal and its leading backgrounds to construct a search for our color-sextet scalar. We in turn use our search to project discovery potential and exclusion limits for this particle after the planned $3\,\text{ab}^{-1}$ run of the (HL-)LHC.

This paper is organized as follows. \hyperref[s2]{Section II} contains a review of color-sextet scalar interactions with color-charged particles and leptons, including the leading production and decay channels at the LHC. In \hyperref[s3]{Section III}, we design a search for these novel particles singly produced and decaying through our dimension-six interactions. In \hyperref[s4]{Section IV}, we compare the sensitivity of our tailored search with that of a search by the CMS Collaboration for $jj\,e^+e^-$ final states from first-generation leptoquark pair production \cite{CMS-EXO-17-009}. We estimate the $3\,\text{ab}^{-1}$ exclusion limits and discovery potential for both searches in \hyperref[s5]{Section V}. Ultimately, we are able to leverage the unique kinematics of our signal --- particularly the distinctive three-body decay of our sextet scalar --- to significantly improve upon the sensitivity of the CMS leptoquark search. In this section we also briefly consider older searches for leptoquarks that are ostensibly capable of excluding sextets somewhat lighter than those covered by LHC. \hyperref[s6]{Section VI} concludes.

%% file: TeX/2_Model.tex
\section{Model discussion}
\label{s2}

We begin by introducing the effective theory of color-sextet scalars to be analyzed in this work. We consider a minimal model of sextet interactions with right-handed Standard Model fermions. If the scalar has weak hypercharge $Y=1/3$, it can couple both to quark pairs of mixed type (one up, one down) and to color-charged particles accompanied by a lepton. Both interactions, however, cannot be realized without inducing lepton number ($L$) non-conservation. We suppose that $L$ is conserved and that the sextet scalar has lepton number $L = -1$, so that its leading interactions with the Standard Model, apart from gauge interactions, are given by
\begin{align}\label{sSmodel}
    \mathcal{L} \supset 
    \frac{1}{\Lambda^2}\,\lambda^{IX}_{u\ell} \bt{J}^{\,s\, ia}\,\Phi_s\,(\,\overbar{u^{\text{c}}_{\text{R}}}_{Ii}\,\sigma^{\mu\nu}\ell_{\text{R}X})\, G_{\mu\nu\,a}
    + \text{H.c.}
\end{align}
with spinor indices contracted within parentheses. The superscript $^{\text{c}}$ denotes charge conjugation and subscripts $_{\text{R}}$ denote right handedness (chirality). The tensor $\sigma^{\mu\nu} = (\ii/2)\,[\gamma^{\mu},\gamma^{\nu}]$ performs a chirality flip. The couplings $\lambda_{u\ell}$ are elements of a matrix in quark and lepton generation space, with $I$ or $X=3$ labeling the heavy generation(s). Finally, the coefficients $\bt{J}$ are the generalized Clebsch-Gordan coefficients \cite{Han:2009ya} required to construct gauge-invariant contractions of the direct-product representation $\boldsymbol{3} \otimes \boldsymbol{6} \otimes \boldsymbol{8}$ in $\mathrm{SU}(3)$ \cite{Carpenter:2021rkl}. These 144 coefficients relate the generators of the $\boldsymbol{6}$ of $\mathrm{SU}(3)$ to the generators of $\boldsymbol{3} \otimes \boldsymbol{8}$ according to
\begin{align}
    [\bt{t}_{\boldsymbol{6}}^a]_s^{\ \ t} &= -\{\bt{J}^{\,s\, ib}\, \bar{\bt{J}}{}_{t\,cj}\, [\bt{t}^a_{\boldsymbol{3}\otimes \boldsymbol{8}}]_{ib}^{\ \ \,jc}\}^*,
\end{align}
with $\bar{\bt{J}}{}_{s\, ai} \equiv [\bt{J}^{\,s\, ia}]^{\dagger}$ denoting Hermitian conjugation, and are normalized according to $\tr \bt{J}^{\,s}\, \bar{\!\bt{J}}_{t}= \delta^s_{\ \, t}$. In the basis where the generators $\bt{t}_{\boldsymbol{3}}$ of the fundamental representation of $\mathrm{SU}(3)$ are proportional to the Gell-Mann matrices, these Clebsch-Gordan coefficients can be written as
\begin{align}
    \bt{J}^{\,s\, ia} = -\ii \epsilon^{ijk} [\bt{t}_{\boldsymbol{3}}^a]_j^{\ \ l} \bar{\bt{K}}{}^s_{\ \,lk},
\end{align}
where $\epsilon^{ijk}$ is the totally antisymmetric symbol and $\bt{K}_s^{\ \,lk}$ are the $lk$-symmetric coefficients for the direct product $\boldsymbol{3} \otimes \boldsymbol{3} \otimes \boldsymbol{\bar{6}}$ of $\mathrm{SU}(3)$, which appear in models of sextet diquarks mentioned in \eqref{diquark} and are tabulated in a few places \cite{Han:2009ya,Han:2010rf}.

For analytic and numerical investigation, we implement this model in \textsc{FeynRules} version 2.3.43 \cite{FR_OG,FR_2}, a package for \textsc{Mathematica}$^\copyright$\ version 12.0 \cite{Mathematica}. We check our analytic results with the help of \textsc{FeynCalc} version 9.3.0 \cite{MERTIG1991345,FC_9.0,FC_9.3.0}, using as input a \textsc{FeynArts} model file generated by \textsc{FeynRules}. Cross sections and event samples are produced at leading order (LO) using \textsc{MadGraph5\texttt{\textunderscore}aMC@NLO} (\textsc{MG5\texttt{\textunderscore}aMC}) version 3.3.1 \cite{MG5,MG5_EW_NLO}, the input for which is a Universal FeynRules Output (UFO) also produced by \textsc{FeynRules} \cite{UFO}. Notes on the UFO implementation of the unique color structure in \eqref{sSmodel}, which is not natively supported by this toolchain, are available in \cite{Carpenter:2021rkl} and in \hyperref[a1]{Appendix A} of this work.

\subsection{Single scalar production}
\label{s2.1}

\begin{figure}
    \centering
    \includegraphics[scale=0.65]{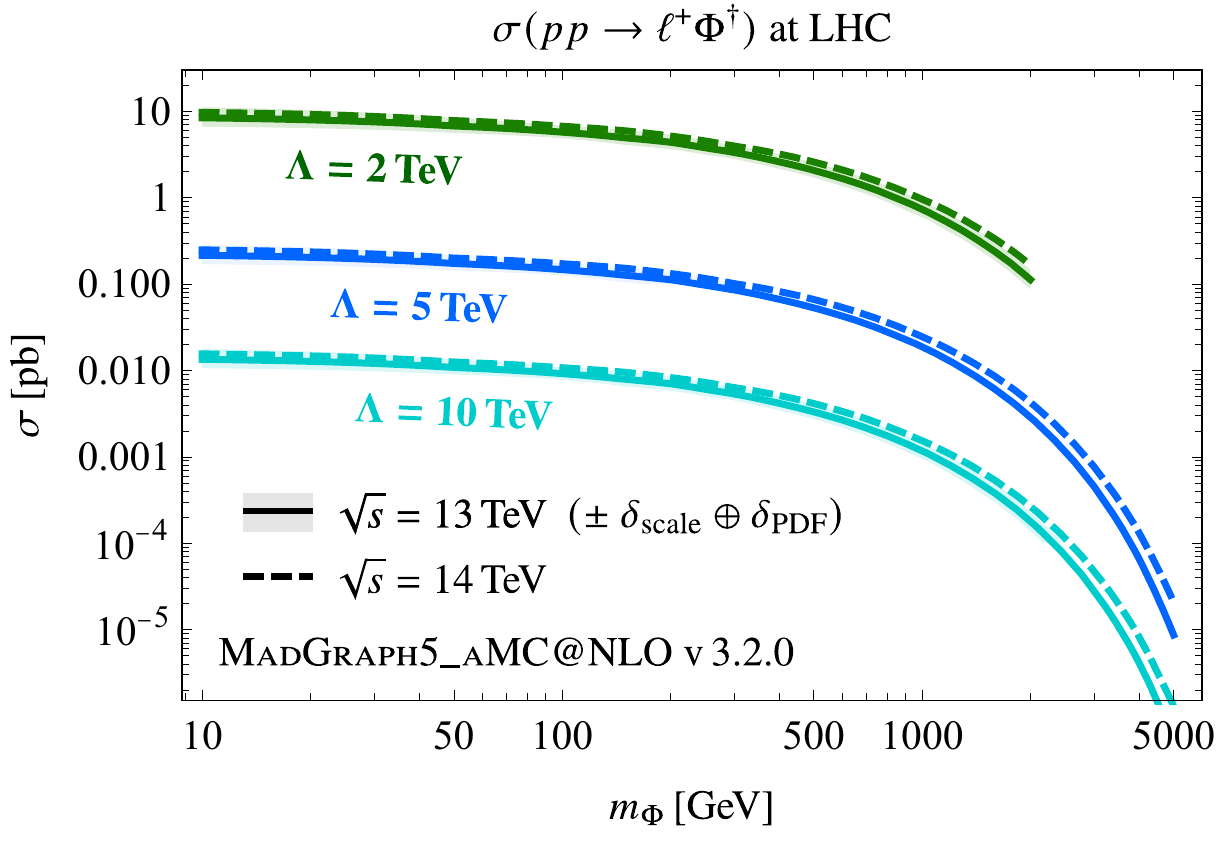}
    \caption{Leading-order inclusive LHC cross sections of color-sextet scalar single production in association with light leptons at $\sqrt{s}=13\,\text{TeV}$ and $\sqrt{s}=14\,\text{TeV}$.}
    \label{xsec1}
\end{figure}

The effective operator \eqref{sSmodel} permits single color-sextet scalar production in association with a lepton \emph{via} quark-gluon fusion at the LHC. Both $\ell^-\Phi$ and $\ell^+\Phi^{\dagger}$ can be produced in this way, but not at equal rates: the latter mode has a $ug$ initial state whose LHC luminosity dwarfs that of the $\bar{u}g$ initial state for $\ell^-\Phi$. We focus on the $\Phi^{\dagger}$ process(es) with larger cross sections. The parton-level cross section for a given initial quark $u_I$ and final-state (anti-)lepton $\ell^+_X$ can be expressed in the massless-quark limit as
\begin{multline}\label{xsecAnalytic}
    \hat{\sigma}(u_I g \to \ell_X^+ \Phi^{\dagger})(\hat{s}) = \frac{1}{64\pi}\left(\frac{\lambda_{u\ell}^{IX}}{\Lambda^2}\right)^2 \frac{1}{\hat{s}}\,(\hat{s}-m_{\Phi}^2 + m_{\ell}^2)\\ \times \left[m_{\ell}^4 - 2m_{\ell}^2(\hat{s}+m_{\Phi}^2) + (\hat{s}-m_{\Phi}^2)^2\right]^{1/2},
\end{multline}
which further collapses to
\begin{align}
    \hat{\sigma}(u_I g \to \ell_X^+ \Phi^{\dagger})(\hat{s}) \approx \frac{1}{64\pi}\left(\frac{\lambda_{u\ell}^{IX}}{\Lambda^2}\right)^2 \frac{1}{\hat{s}}\,(\hat{s}-m_{\Phi}^2)^2
\end{align}
in the massless-lepton limit. In these expressions $\hat{s}$ is the partonic center-of-mass energy. The inclusive LHC cross sections for $\Phi^{\dagger}$ production in association with a light anti-lepton $\ell^+ \in \{e^+,\mu^+\}$ are displayed in \hyperref[xsec1]{Figure 1} for 
\begin{align}\label{benchmark}
    \lambda_{u\ell}^{IX} = 1\ \text{(except for}\ \lambda_{u\ell}^{I3}=\lambda_{u\ell}^{3X}=0),
\end{align}
which we adopt for simplicity throughout the rest of this work, and for the ranges $m_{\Phi} \in [10,5000]\,\text{GeV}$ and $\Lambda \in \{2,5,10\}\,\text{TeV}$.
These cross sections are computed at $\sqrt{s}=13\,\text{TeV}$ and $\sqrt{s}=14\,\text{TeV}$ with renormalization and factorization scales set to the sextet mass $m_{\Phi}$. They are highly sensitive to the cutoff $\Lambda$ of the dimension-six operator \eqref{sSmodel}, with a five-fold increase in $\Lambda$ reducing the cross section from $\mathcal{O}(10)\,\text{pb}$ to a few femtobarns. In this figure, as an initial guess, we take $m_{\Phi} = \Lambda$ as the high end of the range of validity of the effective theory in each benchmark; the $\Lambda=2\,\text{TeV}$ and $\Lambda = 5\,\text{TeV}$ cross sections in \hyperref[xsec1]{Figure 1} are truncated accordingly. Further discussion of the scale of $\Lambda$ is provided in \hyperref[s3]{Section III}, \hyperref[s4]{Section IV}, and (particularly) \hyperref[a2]{Appendix B}.

\subsection{Decays of the sextet scalar}
\label{s2.2}

\begin{figure}
    \centering
    \includegraphics[scale=0.65]{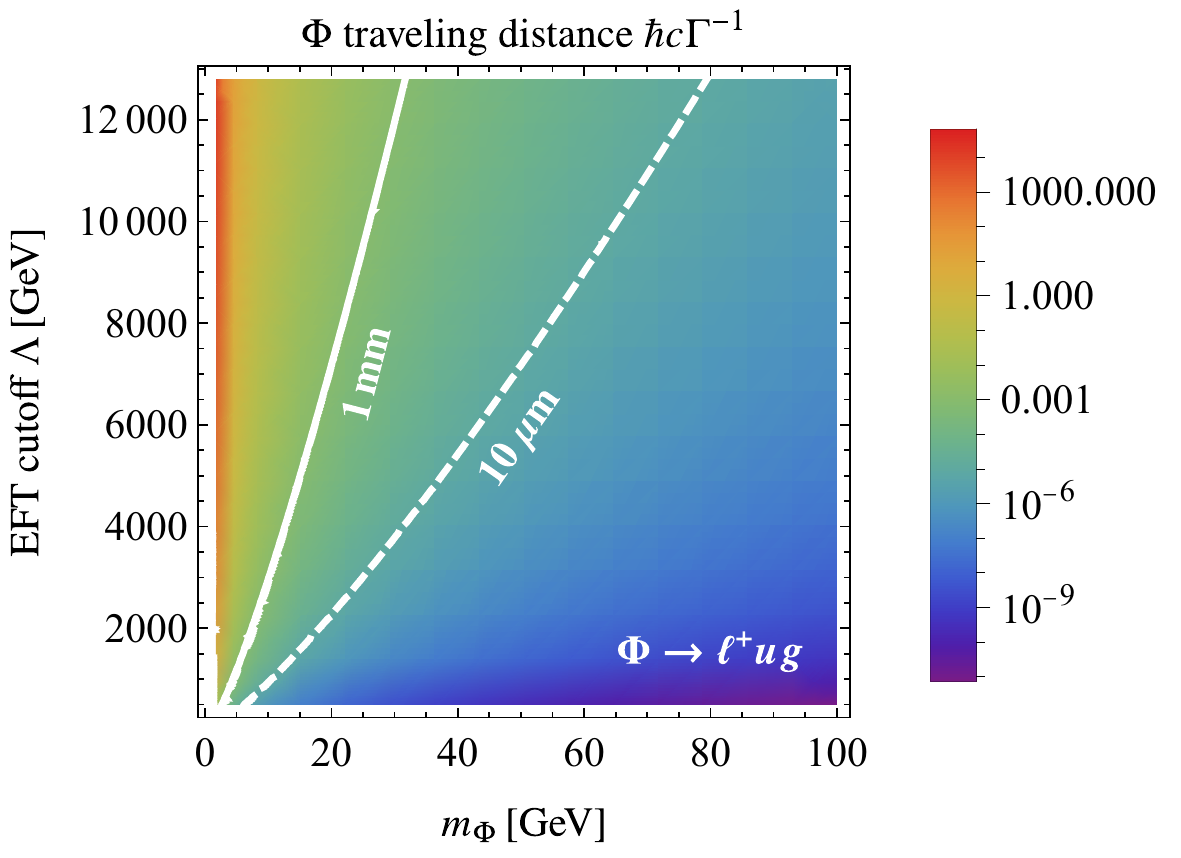}
    \caption{Characteristic traveling distances of a sextet scalar $\Phi$ decaying principally to first- and second-generation SM fermions and a gluon. Contours trace thresholds for displaced vertices at LHC.}
    \label{f1}
\end{figure}

The scalar in our model possesses no two-body decays, decaying only at dimension six to a quark, a lepton, and a gluon (we neglect the four-body decays involving a second gluon). The three-body decay partial width(s) can be expressed as
\begin{multline}\label{3bodywidth}
    \Gamma(\Phi \to \ell^+_X \bar{u}_I g) = \frac{1}{(2\pi)^3}\frac{1}{8m_{\Phi}^3} \left(\frac{\lambda^{IX}_{u\ell}}{\Lambda^2}\right)^2\\ \times \int_{m_{\ell}^2}^{(m_{\Phi}-m_u)^2} \d s_{13}\,\mathcal{F}(s_{13}),
\end{multline}
where $s_{13} = (p_1+p_3)^2$ is the invariant squared mass of the lepton-gluon subsystem (an arbitrary choice), and where
\begin{multline}
    \mathcal{F}(s_{13}) = \frac{1}{s_{13}^2}\,(s_{13}-m_{\ell}^2)^3\,(m_{\Phi}^2 - m_u^2 - s_{13})\\ \times [m_u^4 - 2m_u^2(m_{\Phi}^2 + s_{13}) + (m_{\Phi}^2 - s_{13})^2]^{1/2}
\end{multline}
is proportional to the squared transition amplitude integrated over $s_{23} = (p_2+p_3)^2$. The integral in \eqref{3bodywidth} does not admit an analytical result, except in the massless-fermion limit with $\mathcal{F}(s_{13})\approx s_{13}(m_{\Phi}^2-s_{13})^2$ in which
\begin{align}
\Gamma(\Phi \to \ell^+_X \bar{u}_I g) &\approx 
\frac{2}{3}\frac{1}{(8\pi)^3}\left(\frac{\lambda^{IX}_{u\ell}}{\Lambda^2}\right)^2 m_{\Phi}^5,
\end{align}
but is easy to evaluate numerically in the general case. Before we move on, let us assess whether the sextet scalar should be expected to decay promptly, which will affect how we construct the LHC search below. Following our discussion of sextet production in association with light leptons, suppose that $\Phi$ decays only to first- and second-generation SM fermions. The distance $\hbar c \Gamma^{-1}$ traveled by the scalar before decaying in this manner is displayed in \hyperref[f1]{Figure 2} for the same couplings $\lambda_{u\ell}^{IX}$ in \eqref{benchmark} and a range of light $m_{\Phi}$ and $\Lambda$. We find that suitably light sextets with, for example, $m_{\Phi} \lesssim 35\,\text{GeV}$ and $\Lambda = 5\,\text{TeV}$ may decay after traveling $20\,\mu\text{m}$ or more, possibly producing vertices displaced by a distance greater than the resolution of \emph{e.g.} the CMS detector \cite{email}. More extreme vertex displacement of $\mathcal{O}(1)\,\text{mm}$, well within the regime for long-lived particle searches \cite{ATLAS:2015oan,delaPuente:2015vja}, is possible for $20\,\text{GeV}$ or lighter scalars. This behavior may punch a small hole in the parameter space ostensibly disfavored by searches for promptly decaying resonances. On the other hand, \hyperref[f1]{Figure 2} makes clear that most color-sextet scalar parameter space produces prompt decays and can be targeted with a conventional search strategy. We return to this discussion of unconstrained light-sextet parameter space in \hyperref[s5]{Section V}.

\begin{figure}\label{sig1}
\begin{align*}
\scalebox{0.75}{\begin{tikzpicture}[baseline={([yshift=-.75ex]current bounding box.center)},xshift=12cm]
\begin{feynman}[large]
\vertex (i1);
\vertex [right = 2cm of i1] (i2);
\vertex [above right = 1 cm and 1.25 cm of i2] (v1p);
\vertex [right = 1.5 cm of i2] (v2p);
\vertex [below right = 1 cm and 1.25 cm of i2] (f1);
\vertex [above left= 1.3 cm of i1] (p1);
\vertex [below left = 1.3 cm of i1] (p2);
\vertex [below right = 1.25 cm and 1 cm of i1] (l1);
\diagram* {
(i2) -- [ultra thick, charged scalar] (i1),
(i2) -- [ultra thick, fermion] (v1p),
(i2) -- [ultra thick, fermion,color=blue] (f1),
(v2p) -- [ultra thick, gluon] (i2),
(p1) -- [ultra thick, fermion] (i1),
(p2) -- [ultra thick, gluon] (i1),
(l1) -- [ultra thick, fermion,color=blue] (i1)
};
\end{feynman}
\node at (-1.05,-0.57) {$g$};
\node at (-1.0,0.6) {$u_I$};
\node at (1.2,-0.93) {${\color{blue}\ell^+_X}$};
\node at (1.05,0.375) {$\Phi^{\dagger}$};
\node at (3.4,-0.7) {${\color{blue}\ell^-_Y}$};
\node at (3.5,0.75) {$u_J$};
\node at (3.55,0.325) {$g$};
\node at (0,0) [circle,draw=black,line width = 0.5mm, fill=gray,inner sep=3.5pt]{};
\node at (2,0) [circle,draw=black,line width = 0.5mm, fill=gray,inner sep=3.5pt]{};
\end{tikzpicture}}
\end{align*}
\caption{Diagram for $\Phi^{\dagger}$ production at LHC ($\Phi$ has a $\bar{u}g$ initial state with lower parton luminosity) in association with a light anti-lepton, followed by three-body decay to $\ell u g$. Blobs denote dimension-six effective vertices.}
\end{figure}
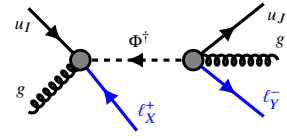

%% file: TeX/3_LHC.tex
\section{A search for sextets with semi-leptonic decays}
\label{s3}

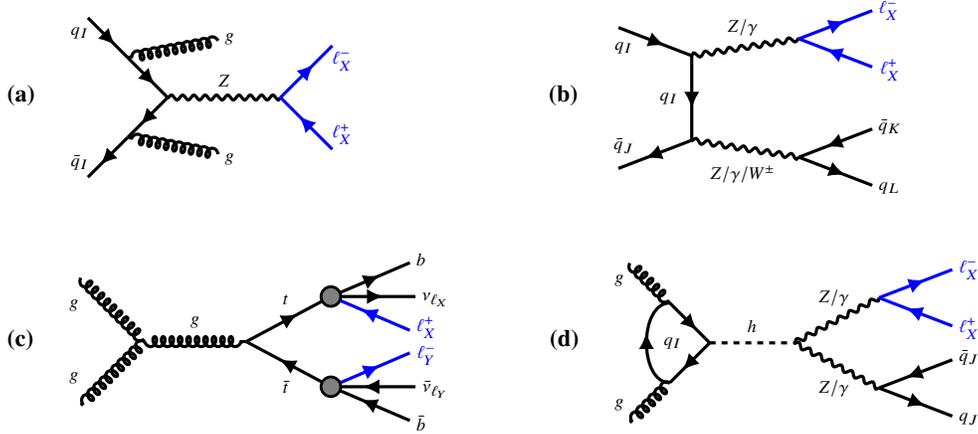
\begin{figure*}\label{bkg1}
\begin{align*}
&\textbf{(a)}\ \ \ \ \ \scalebox{0.75}{\begin{tikzpicture}[baseline={([yshift=-.5ex]current bounding box.center)},xshift=12cm]
\begin{feynman}[large]
\vertex (i1);
\vertex [right = 2cm of i1] (i2);
\vertex [above right = 1.3 cm of i2] (f1);
\vertex [right = 1.5 cm of i2] (v2p);
\vertex [below right = 1.3 cm of i2] (f2);
\vertex [above left= 1 cm of i1] (p1);
\vertex [below left = 1 cm of i1] (g2);
\vertex [above right = 0.3cm and 1.6cm of p1] (r1);
\vertex [below right = 0.3cm and 1.6cm of g2] (r2);
\vertex [above left = 1cm of p1] (q1);
\vertex [below left= 1cm of g2] (q2);
\vertex [below right = 1.25 cm and 1 cm of i1] (l1);
\diagram* {
(q1) -- [ultra thick, fermion] (p1) -- [ultra thick, fermion] (i1) -- [ultra thick, fermion] (g2) -- [ultra thick, fermion] (q2),
(i1) -- [ultra thick, photon] (i2),
(f2) -- [ultra thick, fermion,color=blue] (i2) -- [ultra thick, fermion,color=blue] (f1),
(r2) -- [ultra thick, gluon] (g2),
(r1) -- [ultra thick, gluon] (p1),
};
\end{feynman}
\node at (-1.55,-1.15) {$\bar{q}_I$};
\node at (-1.55,1.15) {$q_I$};
\node at (1.1,-1.1) {$g$};
\node at (1.1,1.05) {$g$};
\node at (1,0.3) {$Z$};
\node at (3.1,-0.65) {${\color{blue}\ell_X^+}$};
\node at (3.1,0.65) {${\color{blue}\ell_X^-}$};
\end{tikzpicture}}\ \ \ \ \ \ \ \ \ \ \ \ \ \ \ \ \ \ \ \ \ \ \ \ \ \ \ \ \ \ \ \ 
\textbf{(b)}\ \ \ \ \ \scalebox{0.75}{\begin{tikzpicture}[baseline={([yshift=-.5ex]current bounding box.center)},xshift=12cm]
\begin{feynman}[large]
\vertex (t1);
\vertex [below=1.5cm of t1] (t2);
\vertex [above left=0.5 cm and 1.3cm of t1] (i1);
\vertex [above right=0.3cm and 1.9cm of t1] (f1);
\vertex [above right=0.5cm and 1.3cm of f1] (p1);
\vertex [below right=0.5cm and 1.3cm of f1] (p3);
\vertex [below left=0.5cm and 1.3cm of t2] (i2);
\vertex [below right = 0.3cm and 1.9cm of t2] (f2);
\vertex [below right=0.5 cm and 1.3cm of f2] (p2);
\vertex [above right=0.5cm and 1.3cm of f2] (p4);
\diagram* {
(i1) -- [ultra thick, fermion] (t1) -- [ultra thick, fermion] (t2) -- [ultra thick, fermion] (i2),
(f1) -- [ultra thick, photon] (t1),
(t2) -- [ultra thick, photon] (f2),
(p3) -- [ultra thick, fermion,color=blue] (f1),
(f1) -- [ultra thick, fermion,color=blue] (p1),
(p4) -- [ultra thick, fermion] (f2),
(f2) -- [ultra thick, fermion] (p2),
};
\end{feynman}
\node at (-1.2,0.15) {$q_I$};
\node at (0.9,0.5) {$Z/\gamma$};
\node at (-1.2,-1.6) {$\bar{q}_J$};
\node at (-0.4,-0.75) {$q_I$};
\node at (0.9,-2.1) {$Z/\gamma/W^{\pm}$};
\node at (3.5, 0.8) {${\color{blue}\ell_X^-}$};
\node at (3.52, -0.25) {${\color{blue}\ell_X^+}$};
\node at (3.52, -1.25) {$\bar{q}_K$};
\node at (3.5, -2.35) {$q_L$};
\end{tikzpicture}}\\[3.5ex]
&\textbf{(c)}\ \ \ \ \ \scalebox{0.75}{\begin{tikzpicture}[baseline={([yshift=-.5ex]current bounding box.center)},xshift=12cm]
\begin{feynman}[large]
\vertex (g1);
\vertex [above left= 1.6 cm of g1] (i1);
\vertex [below left=1.6 cm of g1] (i2);
\vertex [right=1.8cm of g1] (g2);
\vertex [above right= 0.8 cm and 1.5 cm of g2] (t1);
\vertex [above right = 0.6 cm and 1.4 cm of t1] (v1p);
\vertex [right = 1.5 cm of t1] (v2p);
\vertex [below right = 0.6 cm and 1.4 cm of t1] (f1);
\vertex [below right= 0.8 cm and 1.5 cm of g2] (t2);
\vertex [above right = 0.6 cm and 1.4 cm of t2] (v3p);
\vertex [right = 1.5 cm of t2] (v4p);
\vertex [below right = 0.6 cm and 1.4 cm of t2] (f2);
\diagram* {
(i1) -- [ultra thick, gluon] (g1) -- [ultra thick, gluon] (i2),
(g1) -- [ultra thick, gluon] (g2),
(t2) -- [ultra thick, fermion] (g2) -- [ultra thick, fermion] (t1),
(t1) -- [ultra thick, fermion] (v1p),
(t1) -- [ultra thick, fermion] (v2p),
(f1) -- [ultra thick, fermion, color=blue] (t1),
(t2) -- [ultra thick, fermion,color=blue] (v3p),
(v4p) -- [ultra thick, fermion] (t2),
(f2) -- [ultra thick, fermion] (t2),
};
\end{feynman}
\node at (-1.25,0.65) {$g$};
\node at (0.9,0.35) {$g$};
\node at (-1.25,-0.65) {$g$};
\node at (5, 0.25) {${\color{blue}\ell_X^+}$};
\node at (5, -0.25) {${\color{blue}\ell_Y^-}$};
\node at (4.9, 1.45) {$b$};
\node at (4.9,-1.45) {$\bar{b}$};
\node at (5.15,0.75) {$\nu_{\ell_X}$};
\node at (5.15,-0.85) {$\bar{\nu}_{\ell_Y}$};
\node at (2.5, -0.8) {$\bar{t}$};
\node at (2.5, 0.75) {$t$};
\node at (3.3,0.8) [circle,draw=black,line width = 0.5mm, fill=gray,inner sep=3.5pt]{};
\node at (3.3,-0.8) [circle,draw=black,line width = 0.5mm, fill=gray,inner sep=3.5pt]{};
\end{tikzpicture}}\ \ \ \ \ \ \ \ \ \ \ \ \ \ \ \
\textbf{(d)}\ \ \ \ \ \scalebox{0.75}{\begin{tikzpicture}[baseline={([yshift=-.5ex]current bounding box.center)},xshift=12cm]
\begin{feynman}[large]
\vertex (i1);
\vertex [right = 1.5 cm of i1] (i2);
\vertex [above right = 0.8 cm and 1.5 cm of i2] (f1);
\vertex [above right=0.5cm and 1.3cm of f1] (z1);
\vertex [below right=0.5cm and 1.3cm of f1] (z2);
\vertex [right = 1.5 cm of i2] (v2p);
\vertex [below right = 0.8 cm and 1.5 cm of i2] (f2);
\vertex [above right=0.5cm and 1.3cm of f2] (z3);
\vertex [below right=0.5cm and 1.3cm of f2] (z4);
\vertex [above left= 1 cm of i1] (p1);
\vertex [below left = 1 cm of i1] (g2);
\vertex [above right = 0.3cm and 1.6cm of p1] (r1);
\vertex [below right = 0.3cm and 1.6cm of g2] (r2);
\vertex [above left = 1cm of p1] (q1);
\vertex [below left= 1cm of g2] (q2);
\vertex [below right = 1.25 cm and 1 cm of i1] (l1);
\diagram* {
(q1) -- [ultra thick, gluon] (p1) -- [ultra thick, fermion] (i1) -- [ultra thick, fermion] (g2) -- [ultra thick, gluon] (q2),
(i1) -- [ultra thick, scalar] (i2),
(i2) -- [ultra thick, photon] (f1),
(i2) -- [ultra thick, photon] (f2),
(g2) -- [ultra thick, fermion, half left, looseness=1] (p1),
(z2) -- [ultra thick, fermion, color=blue] (f1) -- [ultra thick, fermion,color=blue] (z1),
(z3) -- [ultra thick, fermion] (f2) -- [ultra thick, fermion] (z4),
};
\end{feynman}
\node at (-1.6,-1.1) {$g$};
\node at (-1.6,1.1) {$g$};
\node at (-0.65,0) {$q_I$};
\node at (2.2,-0.8) {$Z/\gamma$};
\node at (2.2,0.8) {$Z/\gamma$};
\node at (0.75,0.3) {$h$};
\node at (4.6,1.3) {${\color{blue}\ell_X^-}$};
\node at (4.6,0.25) {${\color{blue}\ell_X^+}$};
\node at (4.6,-0.25) {$\bar{q}_J$};
\node at (4.6,-1.35) {$q_J$};
\end{tikzpicture}}
\end{align*}
\caption{Representative diagrams for \textbf{(a)} $Z + \text{jets}$ production, \textbf{(b)} diboson production, \textbf{(c)} $t\bar{t}$ production (with $1 \to 3$ decays \emph{via} $W^{\pm}$), and \textbf{(d)} SM Higgs production and decay through $Z/\gamma$, which constitute the leading $jj\, \ell^+\ell^-$ LHC backgrounds for our signal process.}
\end{figure*}

\renewcommand\arraystretch{1.6}
\begin{table*}
    \centering
    \begin{tabular}{l|| S[table-format=3.2] | S[table-format=3.2] | c}
    \toprule
    \hline
     \rule{0pt}{3.5ex}\ \ Background & \ \ \ \ \ \ {$\sigma_{\text{LHC}}^{13\,\text{TeV}}$\,[pb]}\ \ \ \ \ &  \ \ \ \ \ \  {$\sigma_{\text{LHC}}^{14\,\text{TeV}}$\,[pb]}\ \ \ \ \ & Notes \\[1ex]
     \hline
     \hline
\multirow{2}{*}[-0.5ex]{\ \ $Z/\gamma + \text{jets} \to jj\,\ell^+\ell^-$\ \ } & {\multirow{2}{*}[-0.5ex]{\!\!\!1997.43}} & {\multirow{2}{*}[-0.5ex]{\!\!\!2159.63}} & \ \ Up to 2 additional hard jets\ \ \, \\
\cline{4-4}
& & & $m_{\ell\ell} \geq 50\,\text{GeV}$\\
      \hline
    \ \  $ZZ/Z\gamma/\gamma \gamma \to jj\,\ell^+\ell^-$\ \ & 350.2 & 389.1 & \\
      \hline
      \ \  $t\bar{t} \to b\bar{b}\,\ell_X^+\ell_Y^- + \nu_{\ell_X}\bar{\nu}_{\ell_Y}$\ \ & 37.88 & 44.83 & Only process with $E_{\text{T}}^{\text{miss}}$\\
      \hline
    \ \ $W^{\pm}+Z/\gamma \to jj\, \ell^+\ell^-$ & 0.779 & 0.834 & \ \ $Z/\gamma$ decays leptonically\ \ \\
      \hline
    \ \ $h \to ZZ/Z\gamma/\gamma\gamma \to jj\,\ell^+\ell^-$\ \ \ & 0.00893 & 0.0103 & 1 off-shell $Z$ in $h \to ZZ$\\
    \hline
      \bottomrule
    \end{tabular}
    \caption{Leading background processes generating final states with two jets and an opposite-sign lepton pair. Lepton pairs without subscripts have the same flavor. LHC cross sections for $\sqrt{s}=13\,\text{TeV}$ and $\sqrt{s}=14\,\text{TeV}$ are displayed along with event descriptions where helpful.}
    \label{backgroundTable}
\end{table*}
\renewcommand\arraystretch{1}

With this sketch of sextet scalar LHC phenomenology in hand, we now design a search for the single production of these particles. The signal we target in this discussion is represented diagrammatically in \hyperref[sig1]{Figure 3}. It produces final states with two hard jets $j$ and an opposite-sign lepton pair, $\ell^+_X \ell^-_Y$, which need not be of the same flavor \emph{a priori}. On the other hand, most of the largest Standard Model processes with $jj\, \ell^+ \ell^-$ final states produce opposite-sign same-flavor (OSSF) lepton pairs. The leading backgrounds are $\ell^+\ell^-$ production through a photon or $Z$ boson (the Drell-Yan process) with two additional hard jets, $jj\,\ell^+\ell^-$ production through a $Z$ and/or $\gamma$ pair --- produced either non-resonantly or through the decay of a Higgs boson $h$ --- and the similar process with the hadronically decaying $Z/\gamma$ replaced by a $W$ boson. Another process partially overlapping our signal is $t\bar{t}$ production with leptonic decay of the resulting $W^{\pm}$ bosons. This background is unique both because its charged leptons need not share flavor and because two of its final-state leptons are neutrinos, which manifest only as missing transverse energy ($E_{\text{T}}^{\text{miss}}$) in a detector. We provide representative diagrams for these backgrounds in \hyperref[bkg1]{Figure 4} and list in \hyperref[backgroundTable]{Table I} their LHC cross sections at center-of-mass energies of $\sqrt{s}=13\,\text{TeV}$ and $\sqrt{s}=14\,\text{TeV}$ according to the following information.

\subsection{Simulated event samples}
\label{s3.1}

We simulate the Drell-Yan process in \textsc{MG5\texttt{\textunderscore}aMC} version 3.3.1 at LO with up to two additional partons included in the hard-scattering matrix element. The scattering amplitude is convolved with the NNPDF\,2.3 LO set of parton distribution functions \cite{nnpdf} with renormalization and factorization scales $\mu_{\text{R}},\mu_{\text{F}}$ fixed equal to $m_Z$. A generator-level cut of $50\,\text{GeV}$ is placed on the invariant mass $m_{\ell\ell}$ of the decay products of the $Z/\gamma$ to obtain a finite result. The LO hard-scattering events are matched to parton showers in the MLM scheme \cite{Mangano:2006rw} at a scale of 30\,\text{GeV} with the aid of \textsc{Pythia\,8} version 8.244 \cite{Pythia}. We rescale the merged cross section by a next-to-next-to-leading-order (NNLO) enhancement factor ($K$ factor) of $K_{\text{NNLO}} = 1.197$ as computed by \textsc{FEWZ} version 3.1 \cite{Gavin:2010az,Li:2012wna}. This value is compatible with the most recent CMS measurement of the Drell-Yan cross section for $m_{\ell\ell} \geq 50\,\text{GeV}$ \cite{CMS:2018mdl}. Non-resonant pair production of $Z$ bosons decaying semileptonically to $jj\,\ell^+\ell^-$ is simulated at LO and NLO in the strong coupling in \textsc{MG5\texttt{\textunderscore}aMC}, including interference from virtual photons $\gamma^*$. Here again $\mu_{\text{R}}=\mu_{\text{F}}=m_Z$, but for the NLO sample we use the NNPDF\,3.0 NLO parton distribution functions \cite{NNPDF:2014otw}. We shower and hadronize the LO sample for the analysis described below but normalize the sample to the NLO cross section reported by \textsc{MG5\texttt{\textunderscore}aMC}, which exceeds the LO result by a factor of $K_{\text{NLO}} \approx 1.51$. We simulate $W^{\pm} + Z/\gamma$ production in much the same way and find a similar NLO $K$ factor. The $W^+$ processes have slightly larger cross sections at LHC than the $W^-$ analogs, so the total cross section is a bit less than twice the $W^+$ contributions. Top pair production is simulated at LO for a top quark mass of $m_t = 172.5\,\text{GeV}$ with the full leptonic decay chain included in the matrix element. We again use the NNPDF\,2.3 LO parton distribution functions but this time set $\mu_{\text{R}}=\mu_{\text{F}} = m_t$. The normalization of this sample is determined by appropriately combining the branching fractions of leptonic $W^{\pm}$ decays \cite{pdg2020} with the inclusive cross section of $t\bar{t}$ production computed by \textsc{Top++\,2.0} for our choice of $m_t$ at NNLO including resummation of next-to-next-to-leading logarithmic (NNLL) soft-gluon terms \cite{Czakon:2011xx}. Finally, production of a single SM Higgs boson decaying to $ZZ^*$ (or one or more photons) is simulated at LO, which however is at one-loop order \emph{via} gluon fusion (\emph{viz}. \hyperref[bkg1]{Figure 4}; the dominant loop contains a top quark). For this process, the renormalization and factorization scales are fixed equal to the SM Higgs mass. This background has by far the smallest cross section, but in the interest of completeness we rescale the cross section reported by \textsc{MG5\texttt{\textunderscore}aMC} by an NLO $K$ factor of $K_{\text{NLO}} = 1.33$ suitable for the LHC at $\sqrt{s}\approx 13\text{--}14\,\text{TeV}$ \cite{BALL2013746}.\footnote{This $K$ factor varies significantly over a wider range of center-of-mass energies but is fairly flat in the range of current interest.} We simulate all events at both $\sqrt{s}=13\,\text{TeV}$ and $\sqrt{s}=14\,\text{TeV}$; where better information is unavailable, we use the same $K$ factors at both energies. In both cases, the Drell-Yan process is the dominant background, with non-resonant $ZZ$ production non-negligible but subdominant and the other backgrounds small by comparison.
\begin{figure}
    \centering
     \includegraphics[scale=0.85]{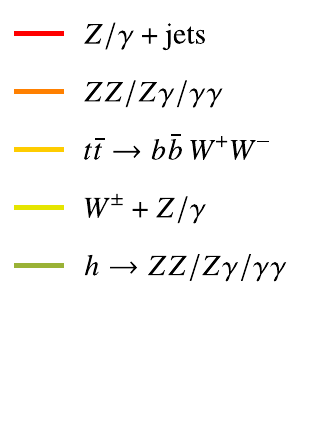}\hspace*{0.5cm}\includegraphics[scale=0.85]{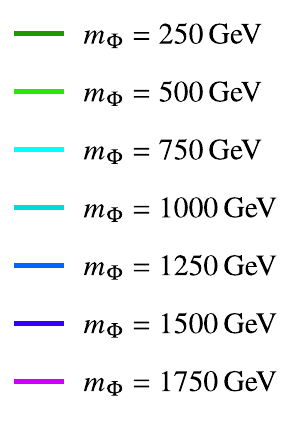}
    \caption{Legend for all distributions displayed in Figures \hyperref[metFig]{6}--\hyperref[reconFig]{9}.}
    \label{legendFig}
\end{figure}

Two background processes that cannot be neglected \emph{a priori} are Higgs boson production in association with a leptonically decaying $Z$ ($hZ \to b\bar{b}\,\ell^+ \ell^-$) and single top quark production in association with a $W$ boson ($tW^- \to b\,\ell^+\ell^- + E_{\text{T}}^{\text{miss}}$ and related), both of which could produce OSSF leptons and multiple jets. While the $hZ$ cross section is small ($\sigma_{\text{LHC}}^{13\,\text{TeV}} = 29.82\,\text{fb}$ \cite{LHC:2016ypw}), the total $tW$ cross section is sizable: $\sigma_{\text{LHC}}^{13\,\text{TeV}} = 71.7\,\text{pb}$ (theory \cite{Kidonakis:2015nna}), $63.1\,\text{pb}$ (ATLAS \cite{ATLAS:2016ofl}), $94\,\text{pb}$ (CMS \cite{CMS:2018amb}). These processes we simulate at LO in \textsc{MG5\texttt{\textunderscore}aMC} and normalize to the aforementioned cross sections (theoretical for $tW$). We find that neither background is kinematically comparable to the leading backgrounds described in previous paragraphs and that, in the end, neither $hZ$ nor $tW$ survives all of the common selection criteria developed over the next few pages to control the leading backgrounds. (In particular, we impose cuts on $E_{\text{T}}^{\text{miss}}$ and the transverse momentum of the hardest jet that fully suppress both processes.) We therefore mention these backgrounds here for completeness but neglect them in the subsequent analysis.
\begin{figure}
    \centering
    \includegraphics[scale=0.6]{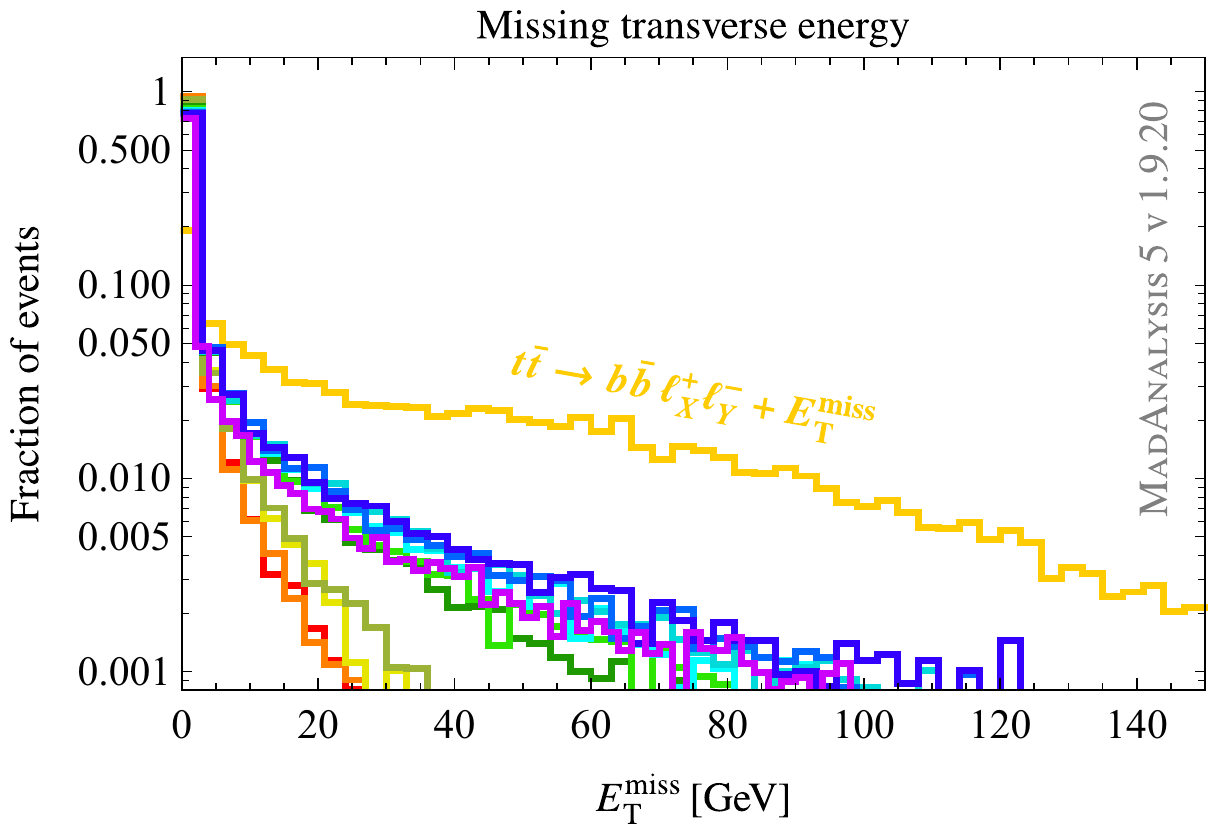}\vspace*{0.75cm}
    \includegraphics[scale=0.6]{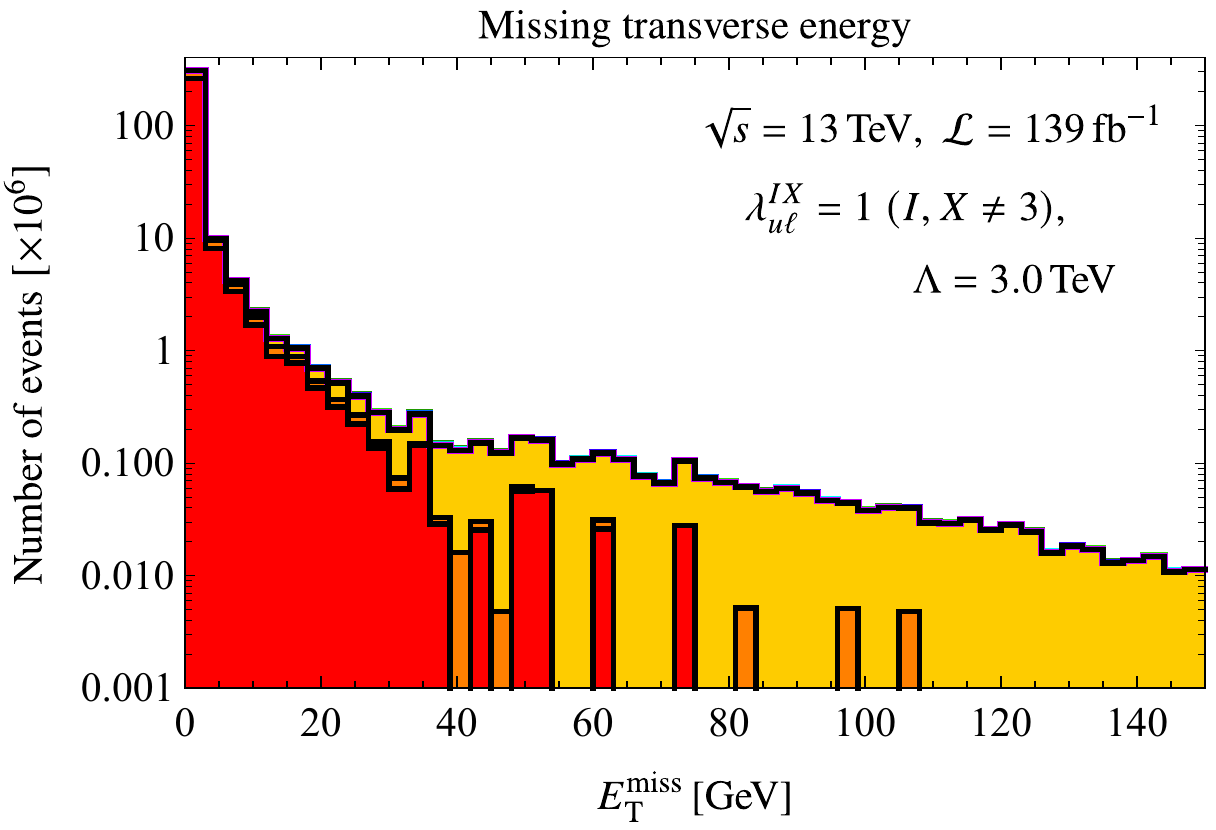}
    \caption{Missing transverse energy ($E_{\text{T}}^{\text{miss}}$) distributions for simulated signals with a range of sextet masses $m_{\Phi}$ and the leading background processes. Full color legend is provided in \hyperref[legendFig]{Figure 5}.}
    \label{metFig}
\end{figure}

\begin{figure*}
    \centering
    \hspace*{0.75cm}\includegraphics[scale=0.6]{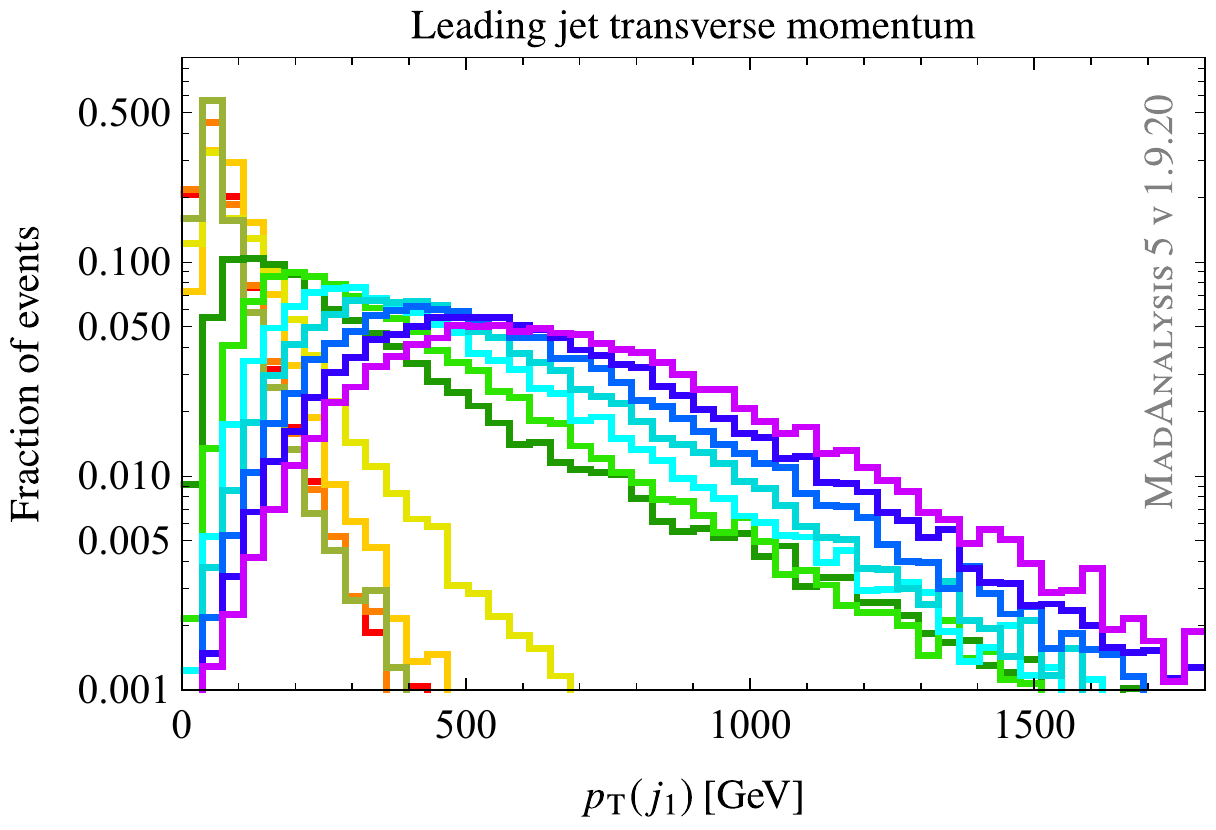}\hfill\includegraphics[scale=0.6]{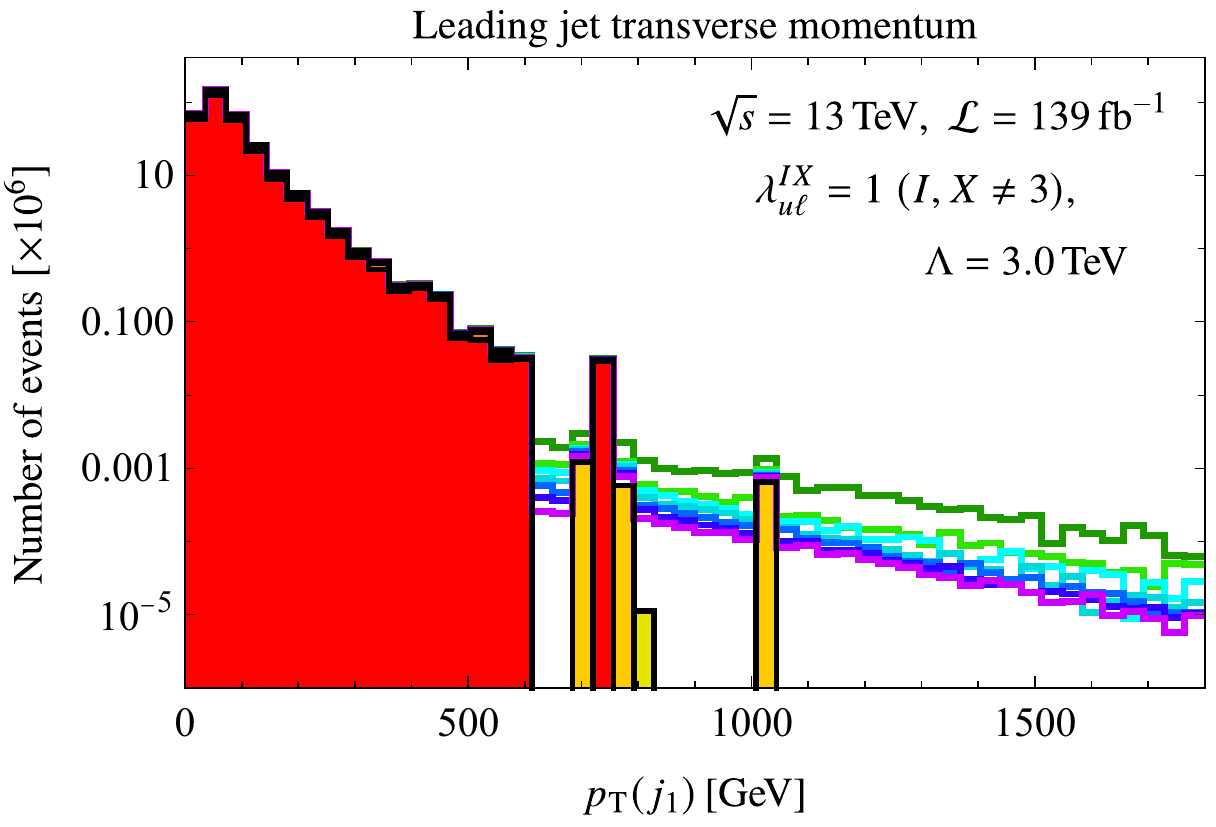}\hspace*{0.75cm}\vspace*{0.75cm}
    \hspace*{0.75cm}\includegraphics[scale=0.6]{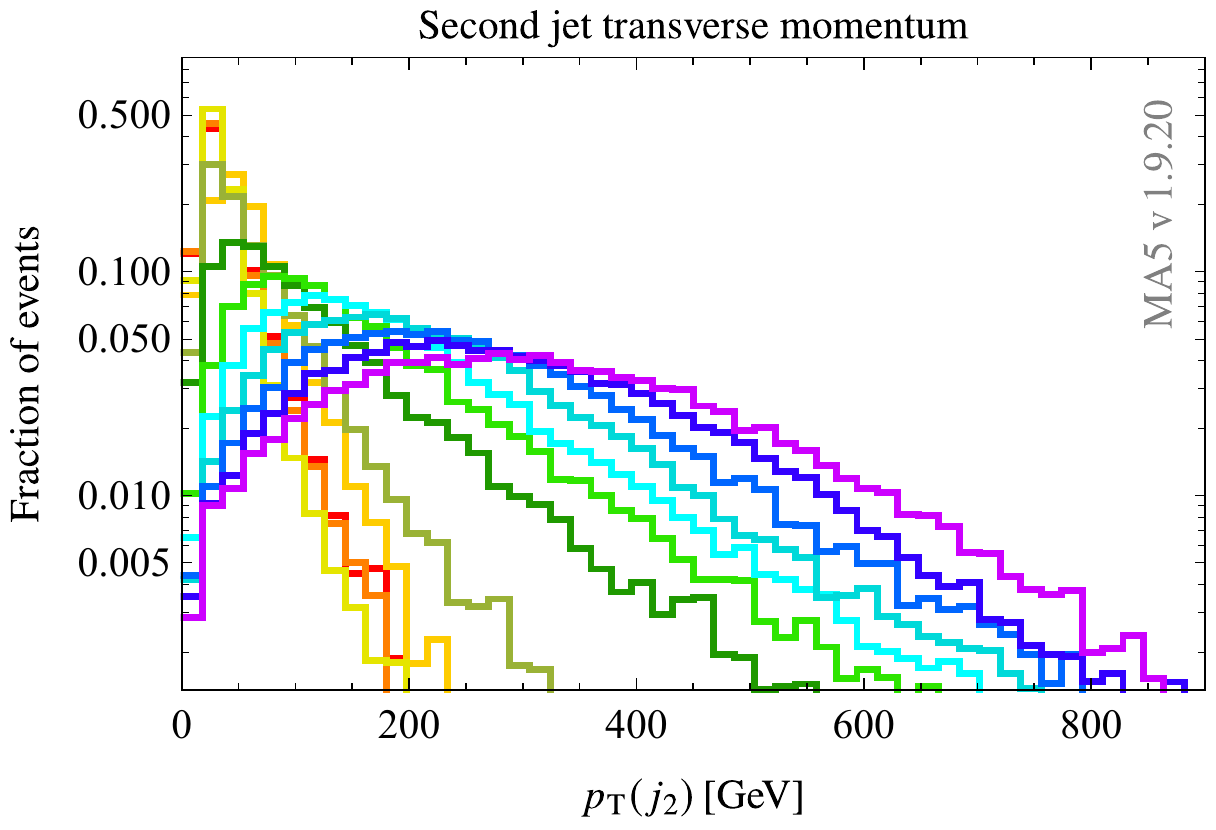}\hfill\includegraphics[scale=0.6]{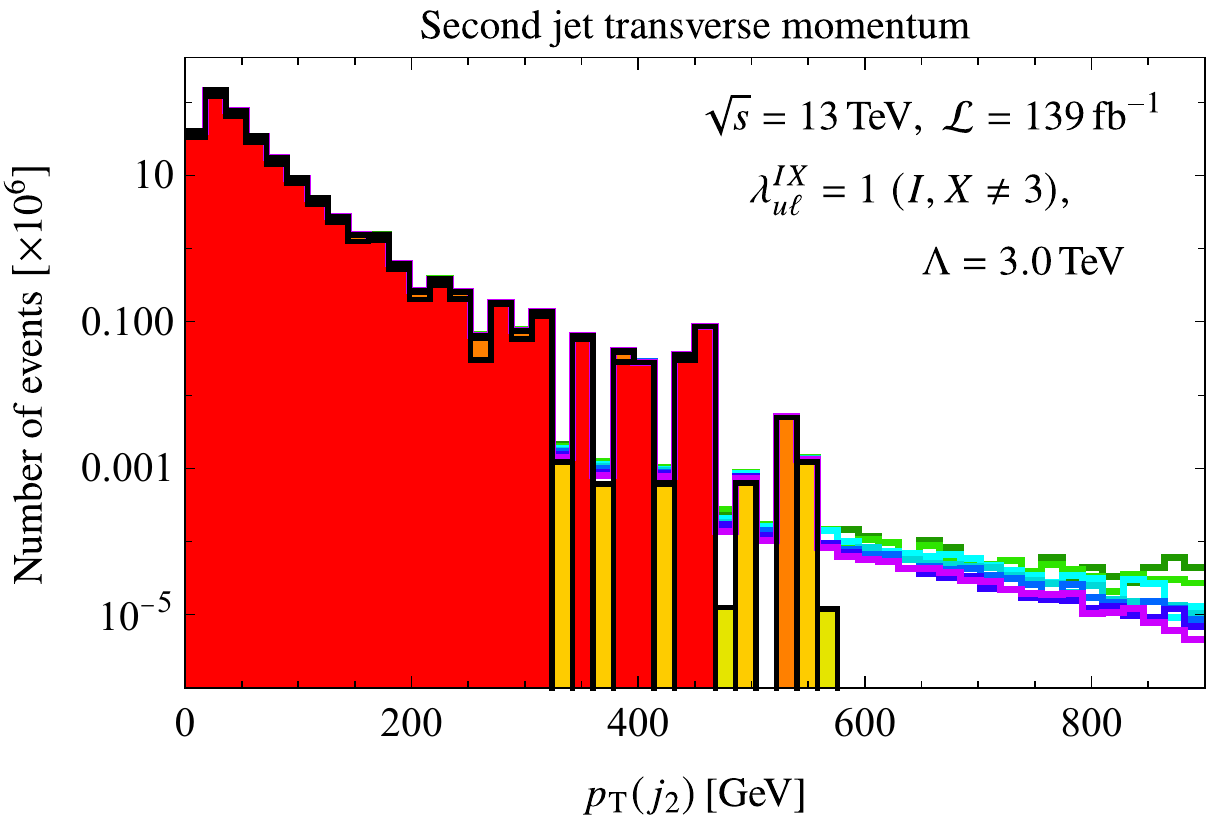}\hspace*{0.75cm}\vspace*{0.75cm}
    \hspace*{0.75cm}\includegraphics[scale=0.6]{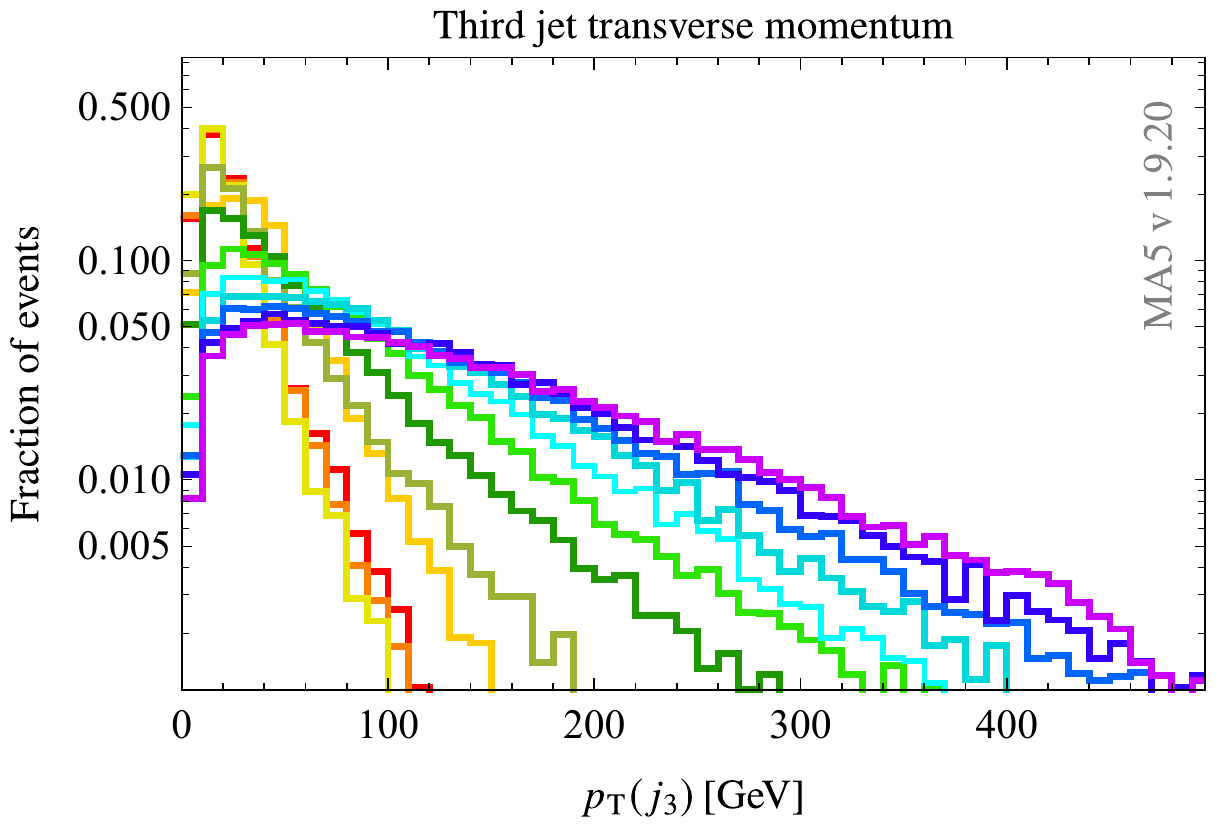}\hfill\includegraphics[scale=0.6]{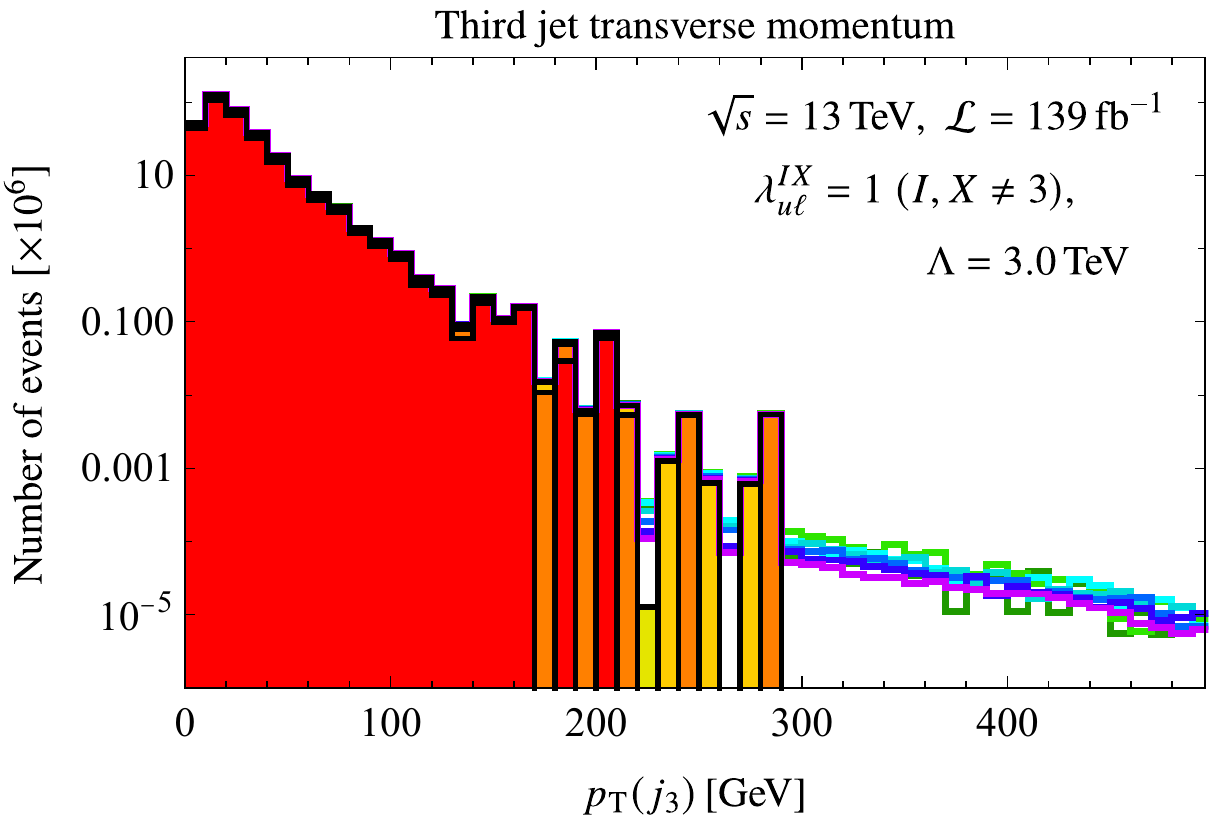}\hspace*{0.75cm}
    \caption{Transverse momentum distributions $p_{\text{T}}(j_1)$, $p_{\text{T}}(j_2)$, $p_{\text{T}}(j_3)$ of the three hardest jets for simulated signals with a range of sextet masses $m_{\Phi}$ and the leading background processes. Left column shows fractions of events and right column shows total number of events at LHC for $\sqrt{s}=13\,\text{TeV}$ and $\mathcal{L} = 139\,\text{fb}^{-1}$. Full color legend is provided in \hyperref[legendFig]{Figure 5}.}
    \label{ptj123Fig}
\end{figure*}

Our first task in designing a search for our semi-leptonically decaying color-sextet scalar is to study the distributions of final-state object multiplicities and kinematics for the signal and background processes. In this exploratory stage, we produce samples of $2.5\times 10^4$ simulated signal events for a range of sextet masses,
\begin{align}
    m_{\Phi} \in \{250,500,750,1000,1250,1500,1750\}\,\text{GeV},
\end{align}
in \textsc{MG5\texttt{\textunderscore}aMC} version 3.3.1, using the UFO input mentioned in \hyperref[s2]{Section II}. Recall that some characteristic cross sections for the processes $pp \to \ell^+\Phi^{\dagger}$ are displayed in \hyperref[xsec1]{Figure 1}. The specific normalization of these samples depends on the branching fractions of $\Phi$ to light leptons --- which we take to be unity; see \hyperref[s2]{Section II} --- and the effective field theory cutoff $\Lambda$ (\emph{viz}. \eqref{sSmodel}). In this discussion we take $\Lambda = 3.0\,\text{TeV}$, a choice motivated not by the ``light-resonance'' bound $\Lambda > m_{\Phi}$ but instead by a more careful partial-wave unitarity analysis provided in \hyperref[a2]{Appendix B}. We combine the corresponding yields with those of the conjugate process $pp \to \ell^- \Phi$ since our search strategy does not distinguish between the two. The background samples discussed above consist of at least the same number of events; a few samples of $3.0 \times 10^4$ events are produced when improved statistical control is required. After being showered with \textsc{Pythia\,8}, all samples are passed to \textsc{MadAnalysis\,5} version 1.9.20 \cite{Conte_2013,Conte_2014,Conte_2018}, which first performs object reconstruction using its inbuilt simplified fast detector simulator (SFS) \cite{Araz_2021} and an interface to \textsc{FastJet} version 3.3.3 \cite{FJ}. The latter reconstructs jets according to the anti-$k_t$ algorithm \cite{Cacciari:2008gp} with radius parameter set to $R=0.4$. The reconstructed samples are finally analyzed by \textsc{MadAnalysis\,5} to produce the distributions shown in Figures \hyperref[metFig]{6}--\hyperref[ptj123Fig]{8}. Every panel in these figures follows the legend displayed at the top of \hyperref[leptonsFig]{Figure 5}.

\subsection{Kinematic distributions and signal significance optimization}
\label{s3.2}

\begin{figure*}
    \centering
    \hspace*{0.75cm}\includegraphics[scale=0.6]{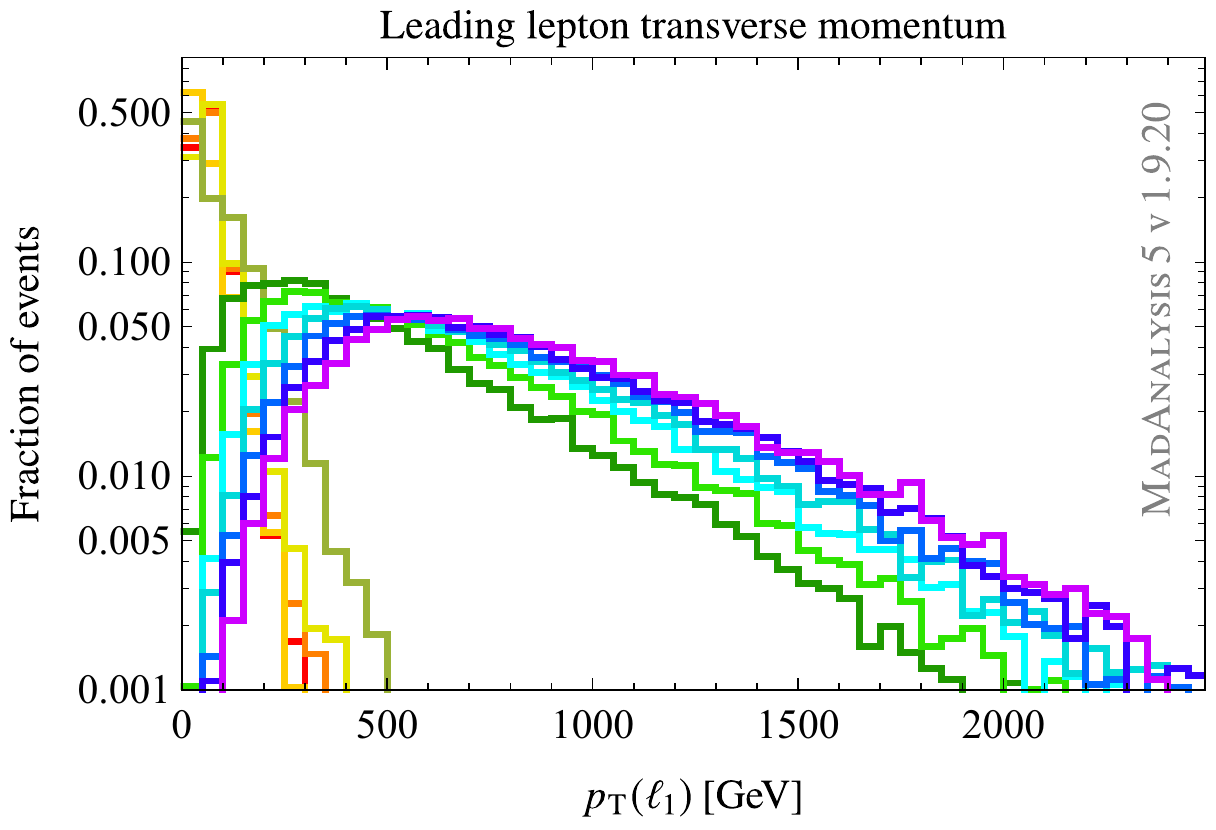}\hfill\includegraphics[scale=0.6]{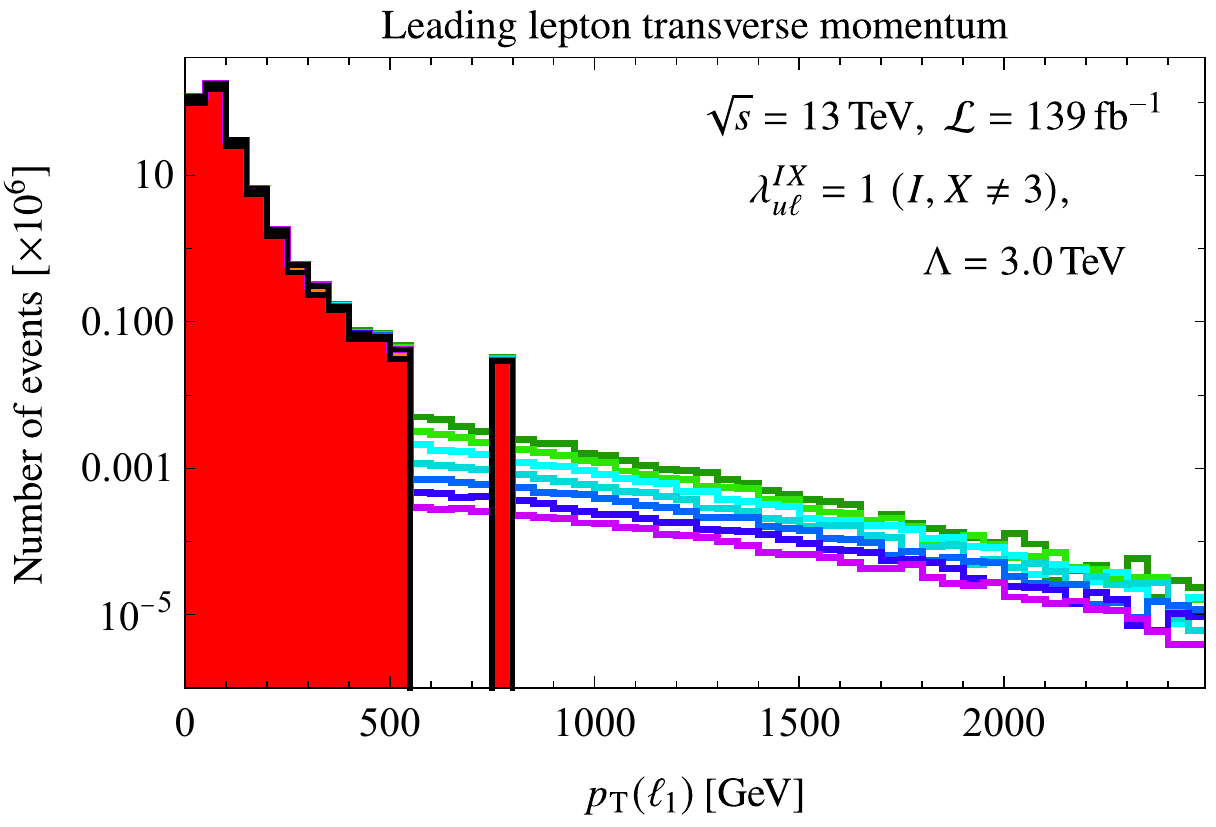}\hspace*{0.75cm}\vspace*{0.75cm}
    \hspace*{0.75cm}\includegraphics[scale=0.6]{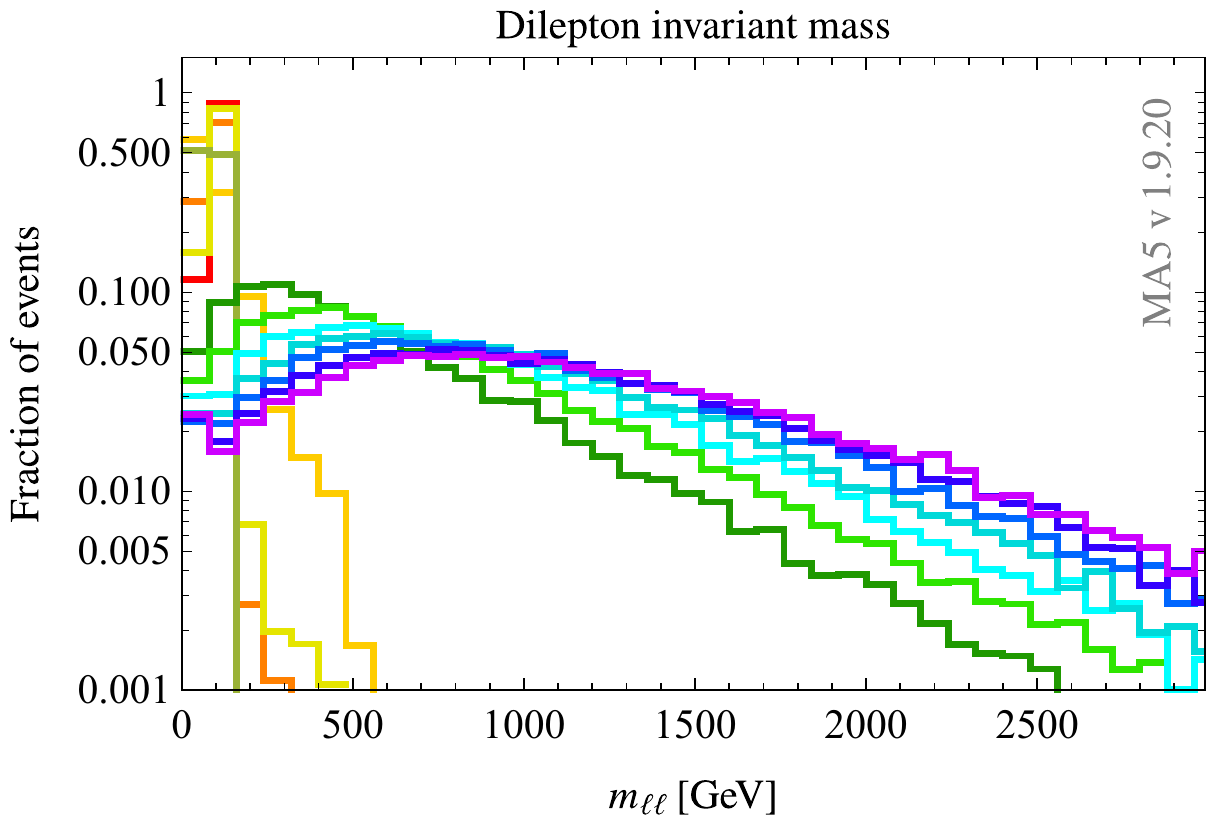}\hfill\includegraphics[scale=0.6]{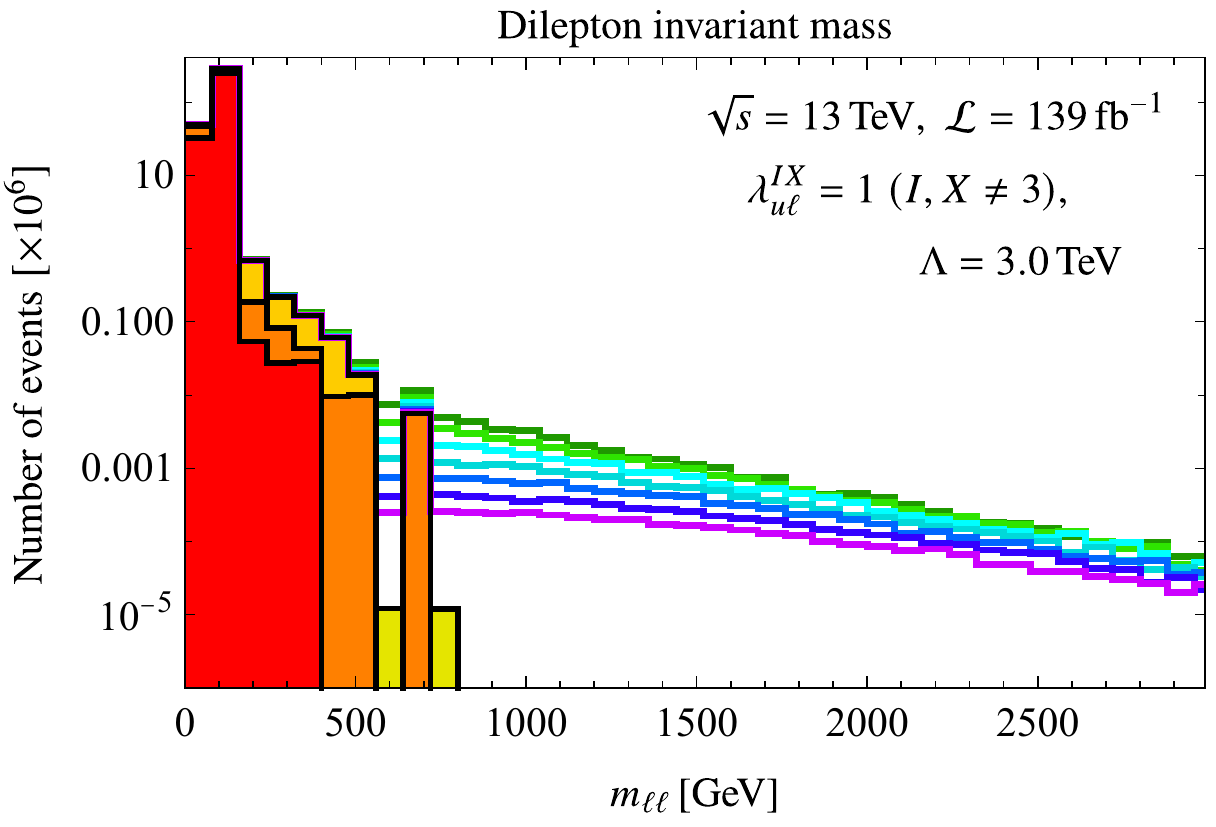}\hspace*{0.75cm}
    \caption{Leading lepton transverse momentum ($p_{\text{T}}(\ell_1)$) and dilepton invariant mass ($m_{\ell\ell}$) distributions for simulated signals with a range of sextet masses $m_{\Phi}$ and the leading background processes. Left column shows fractions of events and right column shows total number of events at LHC for $\sqrt{s}=13\,\text{TeV}$ and $\mathcal{L} = 139\,\text{fb}^{-1}$. Full color legend is provided in \hyperref[legendFig]{Figure 5}.}
    \label{leptonsFig}
\end{figure*}

\begin{figure*}
    \centering
\hspace*{0.86cm}\includegraphics[scale=0.585]{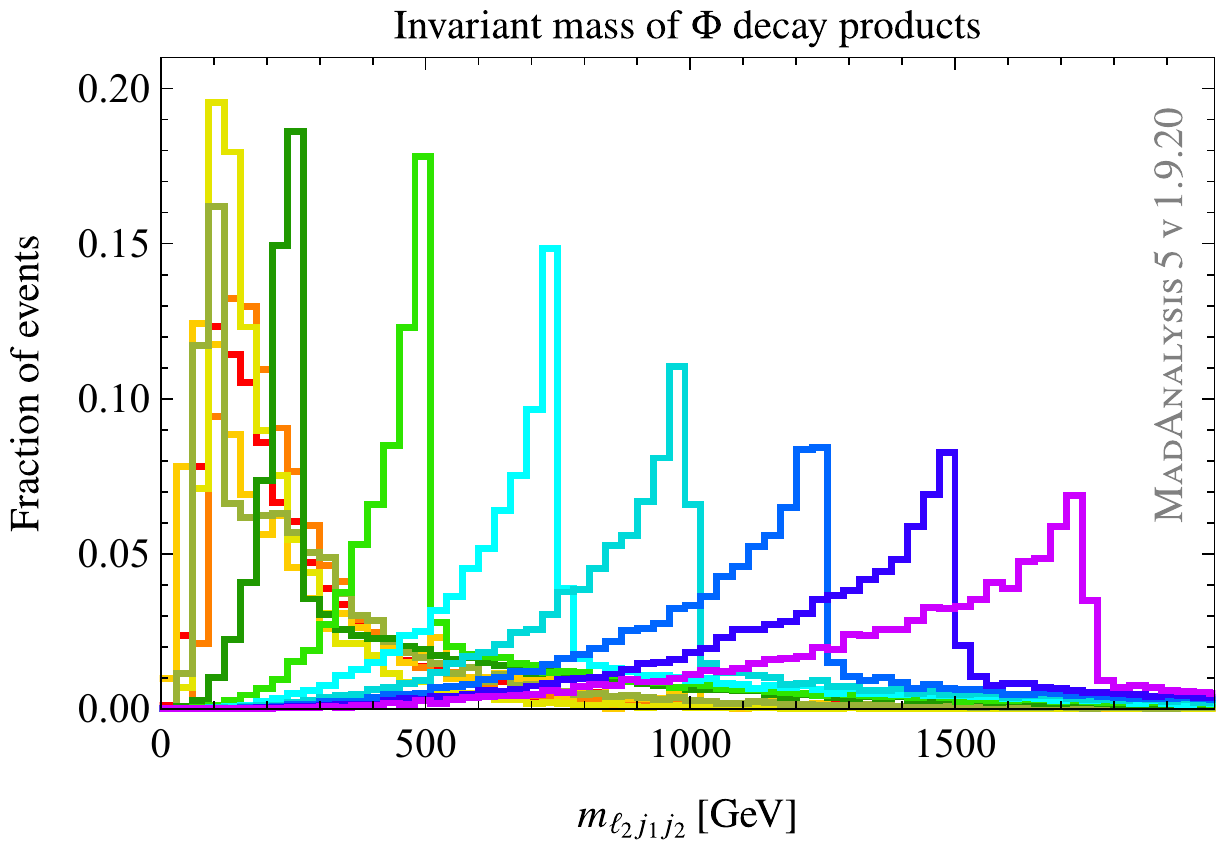}\hfill\includegraphics[scale=0.585]{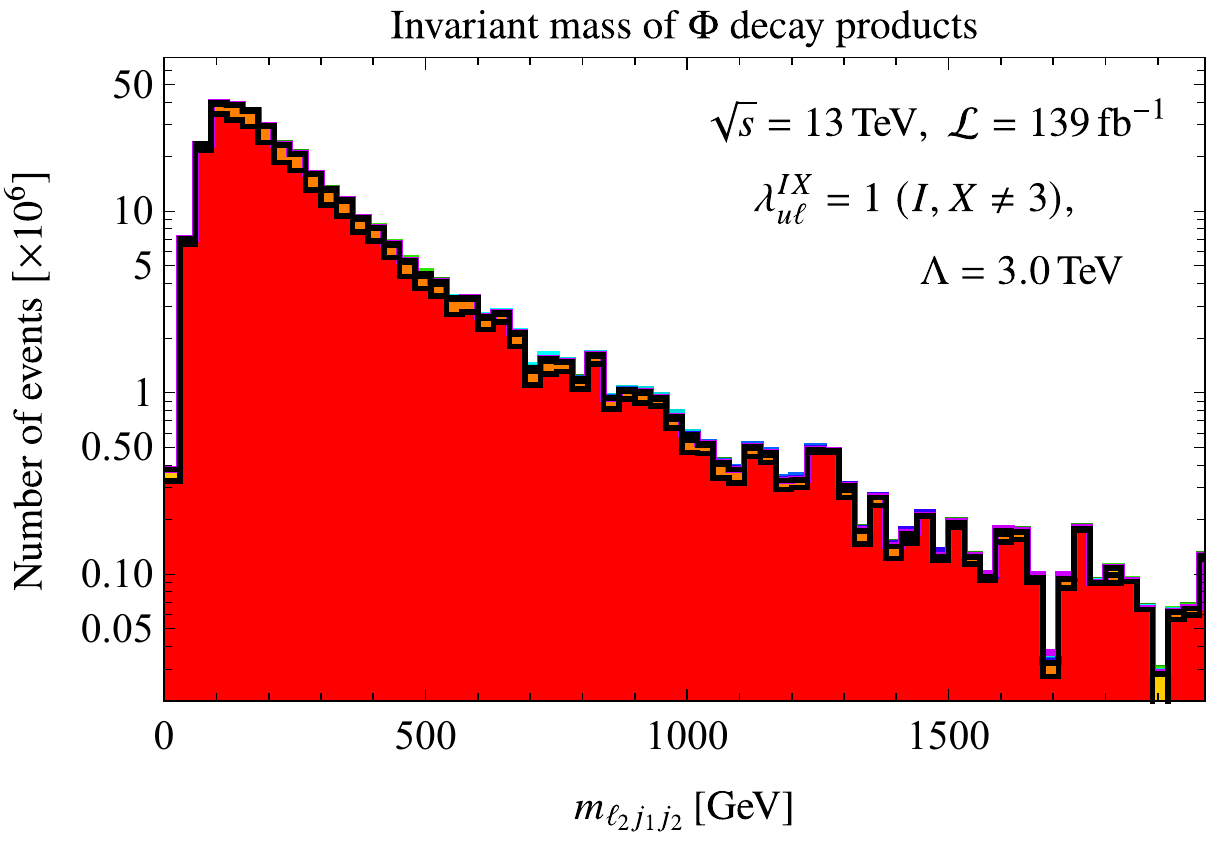}\hspace*{0.75cm}
    \caption{Invariant mass of second lepton and two hardest jets ($m_{\ell_2 j_1j_2}$) distributions for simulated signals with a range of sextet masses $m_{\Phi}$ and the leading background processes. These provide a clean reconstruction of the sextet mass $m_{\Phi}$, but the signals are dwarfed by the Drell-Yan background before selection cuts. Full color legend is provided in \hyperref[legendFig]{Figure 5}.}
    \label{reconFig}
\end{figure*}

\hyperref[metFig]{Figure 6} shows distributions of missing transverse energy $E_{\text{T}}^{\text{miss}}$ for our seven signal samples and the five leading background processes. The top panel shows the fraction of all events with $E_{\text{T}}^{\text{miss}}$ in each of fifty bins in $[0,150]\,\text{GeV}$; \emph{i.e.}, its curves have unit normalization. The lower panel shows (in millions) the total number of events for each process in each bin after the conclusion of LHC Run 2 (hence a center-of-mass energy of $\sqrt{s}=13\,\text{TeV}$ and an integrated luminosity of $\mathcal{L} = 139\,\text{fb}^{-1}$). All signal yields float above the stacked background yields and are not combined with each other. The $13\,\text{TeV}$ background cross sections are displayed in \hyperref[backgroundTable]{Table I}, and we remind the reader that we use signal cross sections corresponding to the sum of $\ell^+\Phi^{\dagger}$ and $\ell^-\Phi$ production in a scenario with $\Lambda= 3.0\,\text{TeV}$ and couplings $\lambda_{u\ell}^{IX}$ as specified by \eqref{benchmark}. The sextet-quark couplings $\lambda_{u\ell}$ in this benchmark are fully specified in the figure. As expected, top pair production ($t\bar{t} \to b\bar{b}\,\ell_X^+ \ell_Y^- + \nu_{\ell_X}\bar{\nu}_{\ell_Y}$) stands out from all other processes since its final-state neutrinos are registered as missing energy. These plots suggest that a stringent cut on maximum $E_{\text{T}}^{\text{miss}}$ may remove most of the $t\bar{t}$ background at little cost to the signals.

\renewcommand\arraystretch{1.6}
\begin{table*}\label{sel1}
\begin{center}
    \begin{tabular}{l|c}
    \toprule
    \hline
\ \ Selection criterion & Selection ranges\\
\hline
\hline
\ \ Jet multiplicity, $N_{\text{jet}}$\ \ \ & $N_{\text{jet}} \geq 2$\\
\hline
\multirow{2}{*}{\ \ Jet kinematics} & $p_{\text{T}}(j) > 15\,\text{GeV}$,\ $|\eta(j)| < 2.5$\\
\cline{2-2}
 & {\color{red}\textsc{veto}}\ \ additional jets with $p_{\text{T}} > 500\,\text{GeV}$\\
 \hline
\ \ Lepton multiplicity & \ 1 opposite-sign (OS) lepton pair\ \ \ \\
\hline
\multirow{4}{*}{\ \ Lepton kinematics} & $p_{\text{T}}(\ell_1) > 20\,\text{GeV}$,\ $p_{\text{T}}(\ell_2) > 15\,\text{GeV}$,\ $|\eta(\ell)| < 2.5$\\
 \cline{2-2}
  & \ \ {\color{red}\textsc{veto}}\ \ additional leptons with $p_{\text{T}} > 10\,\text{GeV}$\ \ \, \\
  \cline{2-2}
 & \ \ {\color{red}\textsc{veto}}\ \ OSSF lepton pairs with $m_{\ell\ell} \in [81,101]\,\text{GeV}$\ \ \, \\
\hline
\ \ Leading lepton isolation\ \ \ & \ \ {\color{red}\textsc{veto}}\ \ any jet with $\Delta R(\ell_1, j) < 0.3$ \\
\hline
\ \ Missing energy, $E_{\text{T}}^{\text{miss}}$\ \ & \ \ {\color{red}\textsc{veto}}\ \ events with $E_{\text{T}}^{\text{miss}} > 25\,\text{GeV}$\\
\hline
\ \ Leading jet momentum & $p_{\text{T}}(j_1) > 350\,\text{GeV}$\\
\hline
\ \ Leading lepton momentum\ \ \ & $p_{\text{T}}(\ell_1) > 300\,\text{GeV}$\\
\hline
\bottomrule
    \end{tabular}
\end{center}
\caption{Common selection criteria of proposed search for color-sextet scalars produced in association with a lepton and decaying to $\ell u g$. OSSF denotes an opposite-sign same-flavor pair resulting from the decay of a $Z$ boson.}
\end{table*}
\renewcommand\arraystretch{1}

\hyperref[ptj123Fig]{Figure 7} shows distributions of transverse momentum $p_{\text{T}}$ for the three leading jets $j_1,j_2,j_3$ in all sample events. In analogy with \hyperref[metFig]{Figure 6}, the left column of \hyperref[ptj123Fig]{Figure 7} shows fractions of all events in each bin while the right column shows the total yields. The benchmark for signal yields is the same as in \hyperref[metFig]{Figure 6} but is again described in the relevant plots. There is a notable separation between the momenta of the two hardest jets in signal and background events that becomes more pronounced for heavier sextets. This behavior is expected since the three-body decay of the scalar, displayed in \hyperref[sig1]{Figure 3}, produces two partons with energies related to the sextet mass $m_{\Phi}$. The comparatively fat tails of these $p_{\text{T}}$ distributions are especially clear in the total-yield plots, which show the background contributions quickly attenuating in the vicinity of $p_{\text{T}}(j) = 500\,\text{GeV}$. The separation between signal and background for the third-hardest jet $j_3$ is not as dramatic, suggesting that cuts on minimum $p_{\text{T}}$ of the two hardest jets may do most of the necessary work in our search.

\renewcommand\arraystretch{1.6}
\begin{table*}\label{cutFlowTable1}
\begin{center}
  \begin{tabular}{l|| c | c | c | c | c }
    \multicolumn{6}{c}{Leading backgrounds, $\sqrt{s} = 13\,\text{TeV}$}\\
    \toprule
    \hline
\ \ Selection criterion \ \ & \ \ $Z + \text{jets}$\ \ \ & \ \ \ \ \ $ZZ$\ \ \ \ \ \ & \ \ \ \ $t\bar{t}$\ \ \ \ \ & \ \ $W^{\pm} + Z$\ \ \ & \ \ $h \to ZZ$\ \ \  \\
\hline
\hline
\ \ $N_{\text{jet}} \geq 2$\ \ & 0.6479 & 0.6847 & \ \ 0.7690\ \ \ & 0.6335 & 0.8249\\
\hline
\ \ 1 OS lepton pair\ \ \ & 0.3409 & 0.5248 & 0.0176 & 0.5266 & 0.4985  \\
\hline
\ \ No OSSF pairs from $Z$\ \ \ & 0.0399 & 0.1086 & 0.0165 & 0.0872 & 0.1334 \\
\hline
\ \ $E_{\text{T}}^{\text{miss}} \leq 25\,\text{GeV}$\ \ \ & 0.0398 & 0.1082 & 0.0015 & 0.0869 & 0.1321\\
\hline
\ \ $p_{\text{T}}(j_1) > 350\,\text{GeV}$\ \ \ & 0.0004 & 0.0005 & 0.0001 & 0.0006 & 0.0042 \\
\hline
\ \ $p_{\text{T}}(\ell_1) > 300\,\text{GeV}$\ \ \ & {\color{FireBrick}0.0000} & 0.0002 & {\color{FireBrick}0.0000} & 0.0002 & 0.0009 \\
\hline
\hline
\ \ $m_{\ell_2j_1j_2} > 200\,\text{GeV}$\ \ \ &  & \ $1.8 \times 10^{-4}$\ \ &  & \ $2.3 \times 10^{-4}$\ \ & \ $8.5 \times 10^{-4}$\ \ \\
\cline{1-1}\cline{3-3}\cline{5-6}
\ \ $m_{\ell_2j_1j_2} > 450\,\text{GeV}$\ \ \ &  & $1.8 \times 10^{-4}$ &  & $2.3 \times 10^{-4}$ & $7.0 \times 10^{-4}$ \\
\cline{1-1}\cline{3-3}\cline{5-6}
\ \ $m_{\ell_2j_1j_2} > 650\,\text{GeV}$\ \ \ &  & $1.1 \times 10^{-4}$ &  & $1.6 \times 10^{-4}$ & $5.8 \times 10^{-4}$ \\
\cline{1-1}\cline{3-3}\cline{5-6}
\ \ $m_{\ell_2j_1j_2} > 900\,\text{GeV}$\ \ \ &  & $6.0 \times 10^{-5}$ & & $1.6 \times 10^{-4}$ & $5.3 \times 10^{-4}$ \\
\cline{1-1}\cline{3-3}\cline{5-6}
\ \ $m_{\ell_2j_1j_2} > 1200\,\text{GeV}$\ \ \ &  & {\color{FireBrick}0.0000} &  & $1.6 \times 10^{-4}$ & $3.2 \times 10^{-4}$ \\
\cline{1-1}\cline{3-3}\cline{5-6}
\ \ $m_{\ell_2j_1j_2} >1400\,\text{GeV}$\ \ \ &  &  &  & $7.7 \times 10^{-5}$ & $2.1 \times 10^{-4}$ \\
\cline{1-1}\cline{5-6}
\ \ $m_{\ell_2j_1j_2} > 1700\,\text{GeV}$\ \ \ &  & &  & {\color{FireBrick}0.0000} & $1.1 \times 10^{-4}$ \\
\cline{1-1}\cline{5-6}
\ \ $m_{\ell_2j_1j_2} > 1900\,\text{GeV}$\ \ \ &  &  &  &  & $1.1 \times 10^{-4}$ \\
\cline{1-1}\cline{6-6}
\ \ $m_{\ell_2j_1j_2} > 2450\,\text{GeV}$\ \ \ &  &  &  &  & $1.1 \times 10^{-4}$ \\
\cline{1-1}\cline{6-6}
\ \ $m_{\ell_2j_1j_2} > 2500\,\text{GeV}$\ \ \ &  &  &  &  & $1.1 \times 10^{-4}$ \\
\hline
\bottomrule
    \end{tabular}
\end{center}
\caption{Cut-flows for background samples in proposed search for color-sextet scalars. Results apply to center-of-mass energy $\sqrt{s} = 13\,\text{TeV}$. Efficiencies are cumulative after each selection cut. Recall that all $Z$ can be replaced by virtual $\gamma$. Jet and lepton multiplicity cuts include kinematic requirements described in \hyperref[sel1]{Table II}. Red entries mark cuts reducing background yields to zero.}
\end{table*}
\renewcommand\arraystretch{1}

\renewcommand\arraystretch{1.6}
\begin{table*}\label{cutFlowTable2}
\begin{center}
\begin{tabular}{l|| c | c | c | c | c | c | c }
    \multicolumn{8}{c}{Light signals, $\sqrt{s}=13\,\text{TeV}$, labeled by $m_{\Phi}$\,[GeV]}\\
    \toprule
    \hline
\ \ Selection criterion \ \ & \ \ \ \ 250\ \ \ \ \ & \ \ \ \ 500\ \ \ \ \ & \ \ \ \ 750\ \ \ \ \ & \ \ \ 1000\ \ \ \ & \ \ \ 1250\ \ \ \ & \ \ \ 1500\ \ \ \ & \ \ \ 1750\ \ \ \ \\
\hline
\hline
\ \ $N_{\text{jet}} \geq 2$\ \ & 0.8914 & 0.9343 & \ \ 0.9501\ \ \ & 0.9591 & 0.9646 & 0.9373 & 0.9763 \\
\hline
\ \ 1 OS lepton pair\ \ \ & 0.7718 & 0.8336 & 0.8558 & 0.8680 & 0.8735 & 0.8780 & 0.8824 \\
\hline
\ \ No OSSF pairs from $Z$\ \ \ & 0.7661 & 0.8303 & 0.8538 & 0.8661 & 0.8720 & 0.8767 & 0.8813 \\
\hline
\ \ $E_{\text{T}}^{\text{miss}} \leq 25\,\text{GeV}$\ \ \ & 0.7349 & 0.7877 & 0.7963 & 0.7986 & 0.8000 & 0.7946 & 0.7932\\
\hline
\ \ $p_{\text{T}}(j_1) > 350\,\text{GeV}$\ \ \ & 0.2525 & 0.3467 & 0.4568 & 0.5539 & 0.6327 & 0.6795 & 0.7145\\
\hline
\ \ $p_{\text{T}}(\ell_1) > 300\,\text{GeV}$\ \ \ & 0.2392 & 0.3301 & 0.4269 & 0.5124 & 0.5945 & 0.6486 & 0.6907\\
\hline
\hline
\ \ {\color{ForestGreen}Best $m_{\ell_2 j_1 j_2}$\,[GeV]}\ \ \ & {\color{ForestGreen}200}  & {\color{ForestGreen}450} & {\color{ForestGreen}650}  & {\color{ForestGreen}900} & {\color{ForestGreen}1200} & {\color{ForestGreen}1400} & {\color{ForestGreen}1700} \\
\hline
\hline
\ \ Final $m_{\ell_2j_1j_2}$ cut\ \ \ & 0.2331 & 0.2745 & 0.3414 & 0.3486 & 0.3953 & 0.3095 & 0.2075 \\
\hline
\bottomrule
    \end{tabular}\\[4ex]
\begin{tabular}{l|| c | c | c | c | c | c | c  }
    \multicolumn{8}{c}{Heavy signals, $\sqrt{s}=13\,\text{TeV}$, labeled by $m_{\Phi}$\,[GeV]}\\
    \toprule
    \hline
\ \ Selection criterion \ \ & \ \ \ 2000 \ \ \ \ & \ \ \ 2500\ \ \ \ & \ \ \ 3000\ \ \ \ & \ \ \ 3500\ \ \ \ & \ \ \ 4000\ \ \ \ & \ \ \ 4500\ \ \ \ & \ \ \ 5000\ \ \ \ \\
\hline
\hline
\ \ $N_{\text{jet}} \geq 2$\ \ & 0.9781 & 0.9818 & \ \ 0.9840\ \ \ & 0.9850 & 0.9855 & 0.9831 & 0.9851 \\
\hline
\ \ 1 OS lepton pair\ \ \ & 0.8839 & 0.8800 & 0.8820 & 0.8855 & 0.8829 & 0.8800 & 0.8800 \\
\hline
\ \ No OSSF pairs from $Z$\ \ \ & 0.8829 & 0.7770 & 0.8820 & 0.8850 & 0.8816 & 0.8800 & 0.8789\\
\hline
\ \ $E_{\text{T}}^{\text{miss}} \leq 25\,\text{GeV}$\ \ \ & 0.7883 & 0.7470 & 0.7664 & 0.7615 & 0.7474 & 0.7422 & 0.7394\\
\hline
\ \ $p_{\text{T}}(j_1) > 350\,\text{GeV}$\ \ \ & 0.7318 & 0.7300 & 0.7500 & 0.7505 & 0.7382 & 0.7333 & 0.7394\\
\hline
\ \ $p_{\text{T}}(\ell_1) > 300\,\text{GeV}$\ \ \ & 0.7114 & 0.7300 & 0.7400 & 0.7430 & 0.7316 & 0.7289 & 0.7349\\
\hline
\hline
\ \ {\color{ForestGreen}Best $m_{\ell_2 j_1 j_2}$\,[GeV]}\ \ \ & {\color{ForestGreen}1900} & {\color{ForestGreen}2450} & \multicolumn{5}{c}{{\color{ForestGreen}2500}} \\
\hline
\hline
\ \ Final $m_{\ell_2j_1j_2}$ cut\ \ \ & 0.1886 & 0.1530 & 0.3700 & 0.4805 & 0.5329 & 0.5540 & 0.5750 \\
\hline
\bottomrule
    \end{tabular}
\end{center}
\caption{Cut-flows for signal samples in proposed search for color-sextet scalars. Results apply to center-of-mass energy $\sqrt{s} = 13\,\text{TeV}$. Efficiencies are cumulative after each selection cut. Jet and lepton multiplicity cuts include kinematic requirements described in \hyperref[sel1]{Table II}.}
\end{table*}
\renewcommand\arraystretch{1}

\hyperref[leptonsFig]{Figure 8} shows distributions of dilepton invariant mass $m_{\ell\ell}$ and the transverse momentum $p_{\text{T}}(\ell_1)$ of the leading lepton $\ell_1$, which is the lepton that recoils off of the sextet scalar at the moment of production ($\ell^+$ for $\Phi^{\dagger}$ production and $\ell^-$ for $\Phi$). The layout and benchmark point for signal yields are the same as in \hyperref[ptj123Fig]{Figure 7}. The $p_{\text{T}}(\ell_1)$ distributions exhibit a striking dichotomy between signals and background, with signal events supporting leptons as hard as $p_{\text{T}} \approx 2.0\,\text{TeV}$. The hard lepton produced in association with the color-sextet scalar may be the most distinctive feature of our signal, and we expect stringent cuts on minimum $p_{\text{T}}(\ell_1)$ to strongly enhance its significance. Meanwhile, there is much more overlap between the signal and background $m_{\ell\ell}$ distributions, and a simple cut on minimum $m_{\ell\ell}$ may reject too many signal events. We therefore restrict ourselves to a veto on invariant masses of OSSF lepton pairs in the vicinity of the $Z$ boson mass in order to cull some events from the large Drell-Yan background, which exhibits a resonance in that area that is visible even with our $m_{\ell\ell}$ resolution and on a logarithmic scale. 

Finally, \hyperref[reconFig]{Figure 9} displays the invariant mass $m_{\ell_2 j_1 j_2}$ of the second-hardest lepton and two hardest jets, which correspond to the decay products of the singly produced color-sextet scalar and whose momenta can therefore be used to reconstruct it. While the right panel shows that the nominal signal yields are much smaller than (in particular) the Drell-Yan background, we see in the left panel that the invariant mass distributions, while asymmetric, are sharply peaked at $m_{\Phi}$ for each displayed sample. We therefore suggest that cutting on the three-body invariant mass could enhance the significance of the sextet signals.

With these distributions in hand, we start to craft a search for color-sextet scalars interacting with light leptons at the LHC. \hyperref[sel1]{Table II} shows common selection criteria intended to define the basic event features and cut away the bulk of the background events without regard for particular color-sextet benchmark scenarios. We require at least two signal jets with $p_{\text{T}} > 15\,\text{GeV}$ located in the central region of either LHC detector ($|\eta| < 2.5$, with $\eta = -\ln \tan\, (\theta/2)$ the pseudorapidity in terms of the polar angle $\theta$ from the beam direction), only two of which are allowed to be very hard ($p_{\text{T}}>500\,\text{GeV}$). We also require a pair of centrally located opposite-sign leptons and forbid OSSF lepton pairs with invariant mass within $10\,\text{GeV}$ of $m_Z \approx 91\,\text{GeV}$. We require the leading lepton to be isolated from all jets by an angular separation of at least $\Delta R(\ell_1,j) = 0.3$, where
\begin{align}\label{deltaR}
\Delta R(X,Y) \equiv \sqrt{(\phi_Y-\phi_X)^2 + (\eta_Y - \eta_X)^2}
\end{align}
is the standard definition in terms of the azimuthal angle $\phi$ around the beam pipe and the pseudorapidity $\eta$ defined above \cite{ATLAS:2012iws}. We finally veto almost all missing transverse energy in order to control the $t\bar{t}$ background as discussed above. These common cuts are followed by some requirement on the minimum invariant mass $m_{\ell_2 j_1 j_2}$. Following our discussion of this observable below \hyperref[reconFig]{Figure 9}, we choose a minimum $m_{\ell_2 j_1j_2}$ in the vicinity of the mass $m_{\Phi}$ of each hypothetical sextet. The asymmetric distributions motivate us to cut $50$--$100\,\text{GeV}$ below each $m_{\Phi}$ up to roughly $m_{\Phi} \sim 2.5\,\text{TeV}$, after which higher $m_{\ell_2 j_1j_2}$ cuts provide negligible additional benefit. The specific choices are made to maximize the signal significance (\emph{viz}. \eqref{significance} in \hyperref[s5]{Section V}) for each simulated $m_{\Phi}$. These final selections are labeled as ``Best $m_{\ell_2 j_1 j_2}$'' in \hyperref[cutFlowTable2]{Table IV}. 

This table, and the preceding \hyperref[cutFlowTable1]{Table III}, respectively display the cumulative signal and background efficiencies under each selection in our search, assuming an LHC center-of-mass energy of $\sqrt{s}=13\,\text{TeV}$. This is done using the same \textsc{MadAnalysis\,5} framework used to study the kinematic distributions. For the backgrounds in \hyperref[cutFlowTable1]{Table III}, we show efficiencies for all $m_{\ell_2 j_1j_2}$ cuts ultimately chosen for the signals in order to track where the background yields are reduced to zero. \hyperref[cutFlowTable2]{Table IV}, meanwhile, shows only the signal efficiency under the $m_{\ell_2 j_1 j_2}$ chosen to maximize the significance. Notably, the large $Z + \text{jets}$ and $t\bar{t}$ backgrounds are eliminated by the final common cut on the leading lepton $p_{\text{T}}$; other significant backgrounds persist for quite a while, and the Higgs background --- while small --- survives all cuts. The signals enjoy efficiencies of 20--60\% under our search, with the final $m_{\ell_2 j_1j_2}$ naturally exhibiting the lowest individual acceptance. We use these results, and their analogs at $\sqrt{s}=14\,\text{TeV}$, to examine the LHC discovery potential in \hyperref[s5]{Section V}.

%% file: TeX/4_CMS-EXO-17-009.tex
\section{Comparison with a search for $\boldsymbol{jj\,e^+e^-}$ final states from leptoquark pair production}
\label{s4}

Before we explore the exclusion and discovery potentials for our color-sextet scalar in the next runs of the LHC, we pause to consider whether existing searches for multiple-jet + dilepton final states are sensitive to our signals. There are a small number of such searches from LHC Run 2 designed to target pair production of scalar leptoquarks (LQs) \cite{BUCHMULLER1987442}, novel $\mathrm{SU}(3)_{\text{c}}$ triplets carrying both baryon and lepton number, which arise not only in unified theories \cite{PhysRevD.8.1240,PhysRevLett.32.438,doi:10.1142/S0217732392000070} but also, increasingly, in phenomenological models seeking to explain possible violations of lepton flavor universality \cite{Dumont:2016xpj,Angelescu:2021lln,Belanger:2021smw}. Since leptoquarks --- by construction --- couple to lepton-quark pairs, the pair production of such states at the LHC is expected to generate $jj\,\ell^+\ell^-$ signatures that may overlap with those considered in \hyperref[s3]{Section III}. On the other hand, the kinematics of these processes are distinct, so it is unclear \emph{a priori} how sensitive LQ-pair searches are to our signals.

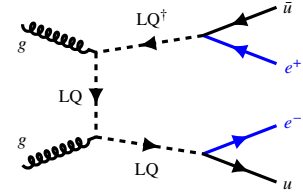
\begin{figure}
    \begin{align*}
        \scalebox{0.75}{\begin{tikzpicture}[baseline={([yshift=-.5ex]current bounding box.center)},xshift=12cm]
\begin{feynman}[large]
\vertex (t1);
\vertex [below=1.5cm of t1] (t2);
\vertex [above left=0.5 cm and 1.3cm of t1] (i1);
\vertex [above right=0.3cm and 1.9cm of t1] (f1);
\vertex [above right=0.5cm and 1.3cm of f1] (p1);
\vertex [below right=0.5cm and 1.3cm of f1] (p3);
\vertex [below left=0.5cm and 1.3cm of t2] (i2);
\vertex [below right = 0.3cm and 1.9cm of t2] (f2);
\vertex [below right=0.5 cm and 1.3cm of f2] (p2);
\vertex [above right=0.5cm and 1.3cm of f2] (p4);
\diagram* {
(i1) -- [ultra thick, gluon] (t1) -- [ultra thick, charged scalar] (t2),
(i2) -- [ultra thick, gluon] (t2),
(f1) -- [ultra thick, charged scalar] (t1),
(t2) -- [ultra thick, charged scalar] (f2),
(p3) -- [ultra thick, fermion,color=blue] (f1),
(p1) -- [ultra thick, fermion] (f1),
(f2) -- [ultra thick, fermion,color=blue] (p4),
(f2) -- [ultra thick, fermion] (p2),
};
\end{feynman}
\node at (-1.3,0.1) {$g$};
\node at (1.0,0.6) {$\text{LQ}^{\dagger}$};
\node at (-1.3,-1.6) {$g$};
\node at (-0.45,-0.8) {$\text{LQ}$};
\node at (0.9,-2.1) {$\text{LQ}$};
\node at (3.4, 0.8) {$\bar{u}$};
\node at (3.5, -0.25) {${\color{blue}e^+}$};
\node at (3.5, -1.25) {${\color{blue}e^-}$};
\node at (3.4, -2.35) {$u$};
\end{tikzpicture}}
    \end{align*}
    \caption{Representative diagram for first-generation leptoquark pair production and decay, which generates a sizable $jj\,e^+e^-$ signal at LHC.}
    \label{LQpairFig}
\end{figure}

Not all leptoquark-pair searches are applicable in principle to our signal. We discard searches for so-called third-generation LQs, which couple only to $t$ or $b$ and $\tau^{\pm}$ \cite{CMS:2020wzx,ATLAS:2021oiz}, and for leptoquarks coupling to light leptons and third-generation quarks \cite{ATLAS:2020dsk}. One Run 2 analysis that appears representative of searches potentially sensitive to our signals is CMS-EXO-17-009, a search by the CMS Collaboration for pair-produced leptoquarks coupling only to electrons and up quarks \cite{CMS-EXO-17-009}. The leptoquark for which this search is optimized has SM quantum numbers $(\boldsymbol{3},\boldsymbol{1},-\tfrac{1}{3})$; its most general renormalizable interactions with first-generation SM fermions are expressed as
\begin{align}\label{LQLag}
    \mathcal{L}_{\text{LQ}} = \Phi_{\text{LQ}}^{\dagger i} \left[y_{\text{L}}\, \overbar{U_{ia}^{\text{c}}}\, \text{P}_{\text{L}}\, \varepsilon^{ab} E_b + y_{\text{R}}\, \overbar{u_{i}^{\text{c}}}\, \text{P}_{\text{R}} e\right] + \text{H.c.},
\end{align}
where $\{U,E\}$ are the first-generation quark and lepton $\mathrm{SU}(2)_{\text{L}}$ doublets with indices $a,b \in \{1,2\}$, and $\{u,e\}$ are the corresponding weak singlets. (Otherwise we follow the conventions of \eqref{sSmodel}.) A representative diagram for the pair production and fully visible decay $(\text{LQ} \to e^- u$) of these first-generation LQs is displayed in \hyperref[LQpairFig]{Figure 10}, and other pair-production diagrams can be found in Figure 1 of CMS-EXO-17-009.

The analysis looks for these leptoquarks using $\mathcal{L} = 35.9\,\text{fb}^{-1}$ of $pp$ collisions at $\sqrt{s}=13\,\text{TeV}$. There are two sub-analyses: one is in the ``$eejj$'' channel represented in \hyperref[LQpairFig]{Figure 10}, and the other is in a channel where one LQ decays to an electron neutrino and a down quark (``$e\nu jj$''), which is permitted by the left-chiral term in \eqref{LQLag}. Only the $eejj$ search overlaps with our sextet signal, so we set aside the $e\nu jj$ sub-analysis. The $eejj$ baseline selection requires at least two isolated electrons, with no stipulations on their electric charges, and at least two central ($|\eta| < 2.4$) jets, all with transverse momentum $p_{\text{T}} > 50\,\text{GeV}$. A veto is imposed on muons with $p_{\text{T}} > 35\,\text{GeV}$ and $|\eta|<2.4$, which are however used to define a control region employed to estimate the $t\bar{t}$ background (the very same process discussed in \hyperref[s3]{Section III} for our signal). This veto obviously renders this search insensitive to color-sextet scalars in our model decaying to muons. The baseline selection is supplemented by common selection criteria for the leading electrons and jets. In particular, the invariant mass $m_{ee}$ of the leading electron pair must exceed $50\,\text{GeV}$, the transverse momentum of this system must be greater than $70\,\text{GeV}$, and the scalar $p_{\text{T}}$ sum
\begin{align}\label{STdef}
S_{\text{T}} = p_{\text{T}}(e_1) + p_{\text{T}}(e_2) + p_{\text{T}}(j_1)+ p_{\text{T}}(j_2)
\end{align}
over the same electrons and the two leading jets must exceed $300\,\text{GeV}$. The $eejj$ search is then finalized with selection criteria tailored to a range of LQ masses; namely,
\begin{align}\label{CMSbins}
    m_{\text{LQ}} \in [200,1050]\,\text{GeV}\ \ \ \text{with}\ \ \ \Delta m_{\text{LQ}} = 50\,\text{GeV}.
\end{align}
The final selections are displayed in Figure 2 of CMS-EXO-17-009 and are imposed on $m_{ee}$, $S_{\text{T}}$, and $m_{ej}^{\text{min}}$, the smaller of the two electron-jet invariant masses computed after choosing the pairings of the two leading electrons and jets such that the difference between the invariant masses is minimized. The $m_{\text{LQ}} = 1.05\,\text{TeV}$ selections are used for all hypothetical leptoquarks with mass in excess of that figure.

\begin{figure}
    \centering
    \includegraphics[scale=0.6]{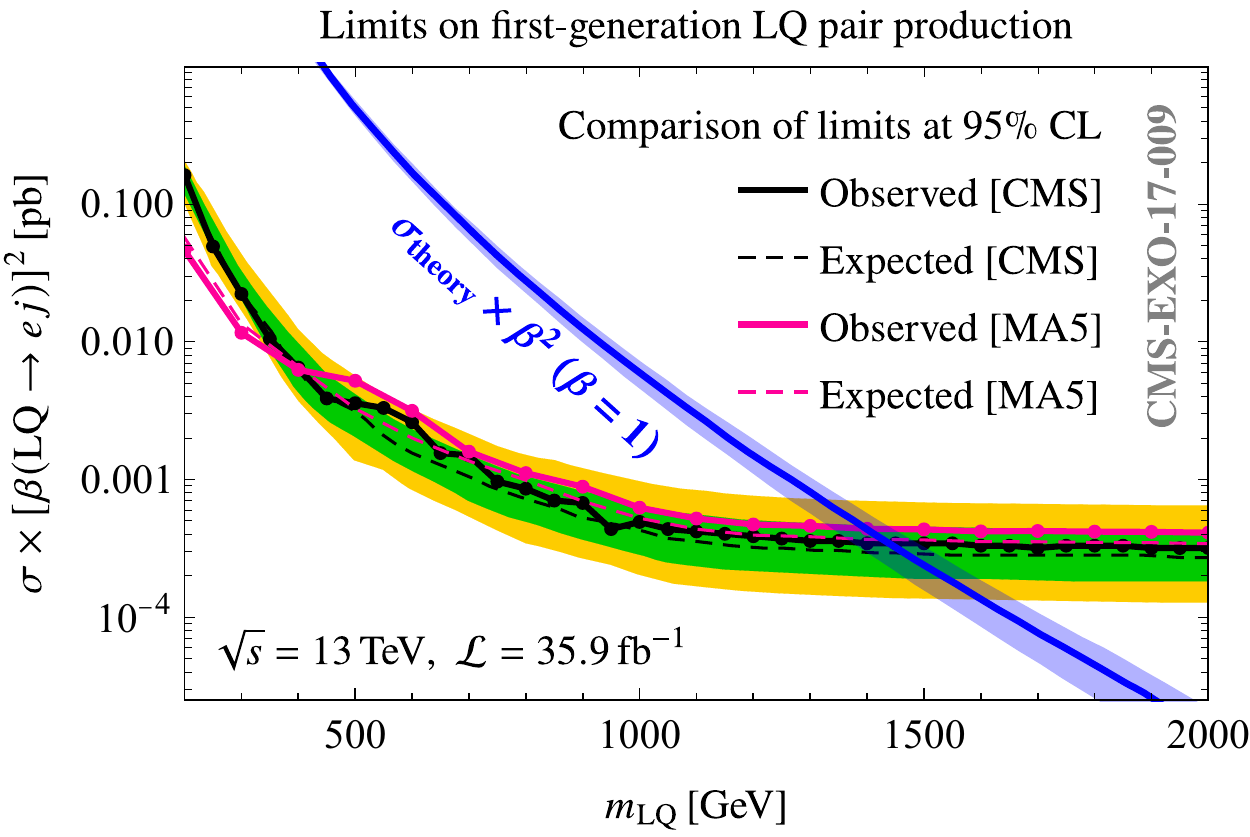}
    \caption{Observed and expected limits at 95\% CL obtained with \textsc{MadAnalysis\,5} implementation of CMS-EXO-17-009, a search for first-generation leptoquark pair production in $jj\,e^+e^-$ final states, compared to official CMS limits.}
    \label{LQrecastFig}
\end{figure}

CMS finds no significant deviation from the SM expectation and computes upper limits at 95\% confidence level (CL) \cite{Read:2002cls} on the cross section of LQ pair production followed by decay to $ej$. The observed and expected limits are provided in Figure 8 of the analysis and are reproduced in our \hyperref[LQrecastFig]{Figure 11}.
The efficiencies under the full $eejj$ selection of the LQ signal simulated by CMS are helpfully provided in Figure 11 of the analysis. These results, alongside the detailed description of the selection criteria, provide enough information for us to recast the analysis in \textsc{MadAnalysis\,5}. We implement the CMS $eejj$ selection in the expert mode of \textsc{MadAnalysis\,5} and link our code to \textsc{Delphes\,3} version 3.4.2 for simulation of the CMS detector response and \textsc{FastJet} version 3.3.3 for object reconstruction \cite{Delphes_3}. 

\begin{figure}
    \centering
    \includegraphics[scale=0.85]{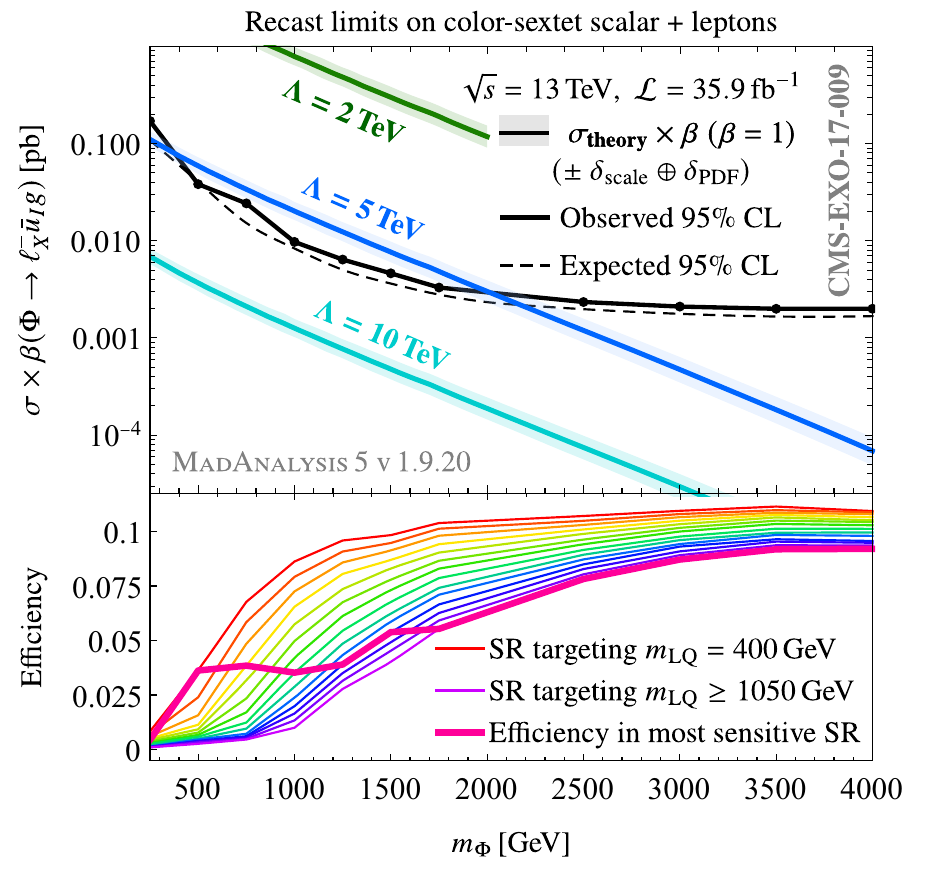}
    \caption{Efficiencies (bottom) and observed and expected limits at 95\% CL (top) imposed on sextet scalar production in association with light leptons by CMS-EXO-17-009.}
    \label{recastResultsFig}
\end{figure}

In order to validate our implementation, we produce samples of $2.5 \times 10^4$ leptoquark signal events in close correspondence with the simulated signals described in CMS-EXO-17-009. We implement the model, led by the interactions \eqref{LQLag}, in \textsc{FeynRules} version 2.3.43 and use \textsc{NLOCT} version 1.02 \cite{NLOCT} to automatically renormalize the model and produce a UFO module capable of NLO QCD calculations. We compute the pair-production cross sections for a range of LQ masses $m_{\text{LQ}}$ at NLO in \textsc{MG5\texttt{\textunderscore}aMC} version 3.3.1, finding good agreement with the results quoted by CMS \cite{Kramer:2004df} after setting the renormalization and factorization scales fixed to each $m_{\text{LQ}}$ and using the CTEQ6L1 set of parton distribution functions \cite{Pumplin:2002vw}. However, in accordance with the experimental analysis, we simulate the full events including $ej$ decay at LO, using the NNPDF\,2.3 LO parton distribution functions, and use the CMS normalizations displayed in \hyperref[LQrecastFig]{Figure 11}. We shower and hadronize the events using \textsc{Pythia\,8} version 8.244 and use these showered samples as input for the reconstruction mode of \textsc{MadAnalysis\,5}. The resulting observed and expected limits at 95\% CL are displayed in \hyperref[LQrecastFig]{Figure 11} alongside the official results. We find good agreement between the results, especially in the region where the signal is excluded, where the discrepancy between excluded $m_{\text{LQ}}$ is roughly $50\,\text{GeV}$ or $3.5\%$.

\begin{figure*}
    \centering
\hspace*{0.75cm}\includegraphics[scale=0.6]{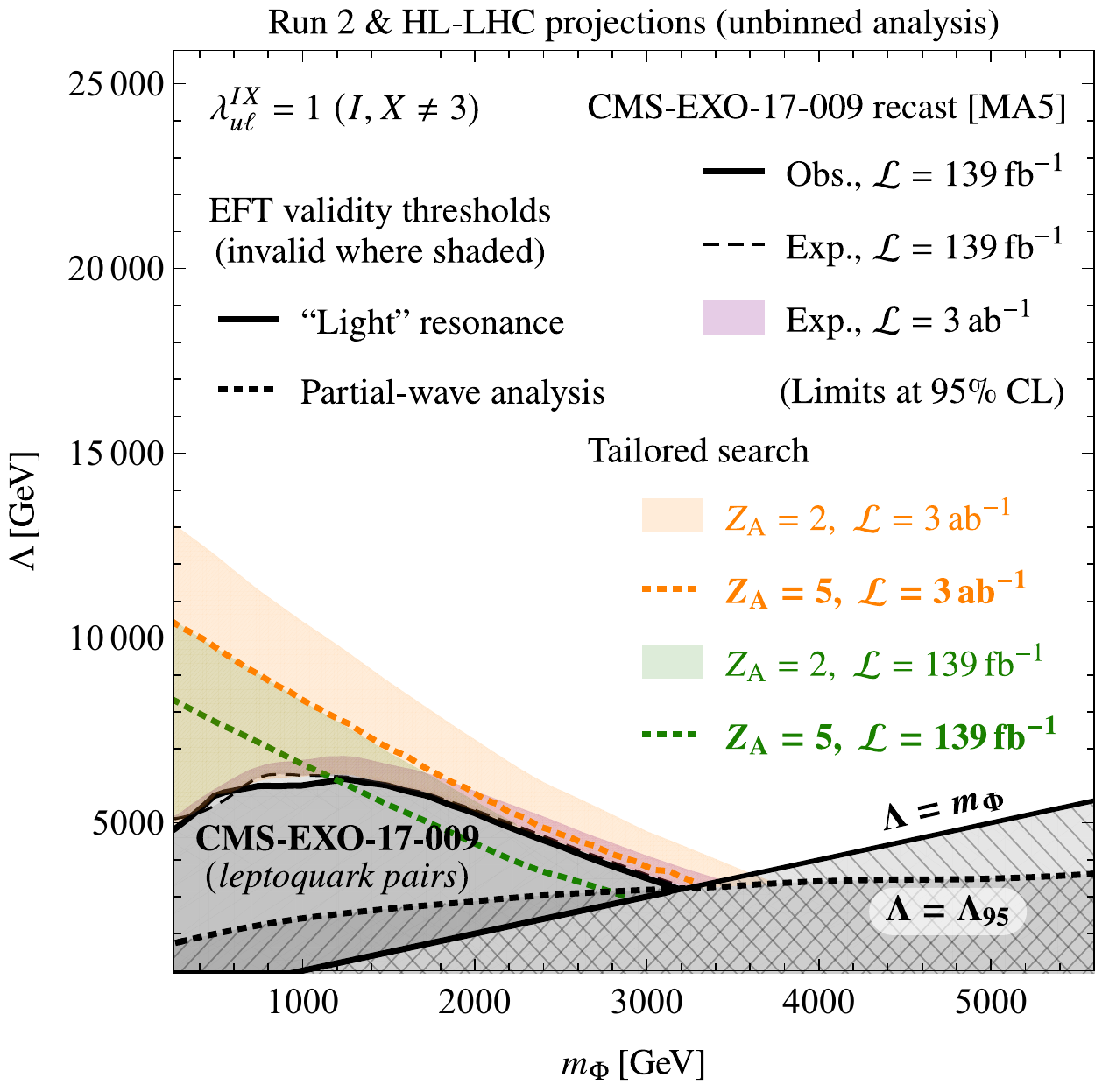}\hfill\includegraphics[scale=0.6]{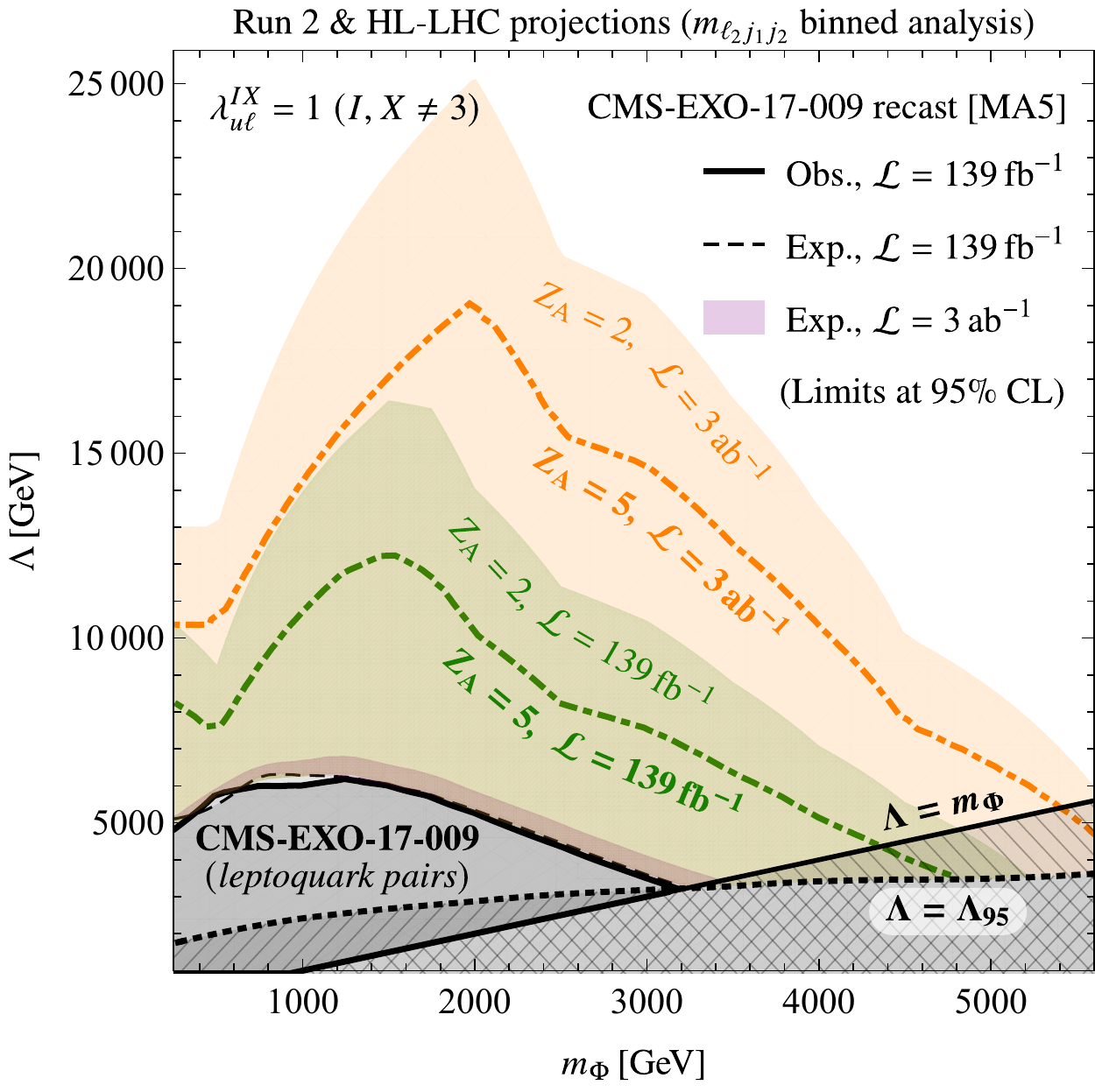}\hspace*{0.75cm}
    \caption{Projected contours of discovery ($Z_{\text{A}} = 5$) and exclusion in the absence of a signal ($Z_{\text{A}} = 2$) in our EFT parameter space for samples corresponding to the full LHC Run 2 dataset ($\mathcal{L} = 139\,\text{fb}^{-1}$, $\sqrt{s}=13\,\text{TeV}$) and the planned HL-LHC dataset ($\mathcal{L} = 3\,\text{ab}^{-1}$, $\sqrt{s}=14\,\text{TeV}$). Left panel shows results from unbinned search; \emph{i.e.}, before $m_{\ell_2 j_1j_2}$ selections; right panel shows sensitivity gains from final $m_{\ell_2 j_1j_2}$ cuts. Projected limits from CMS-EXO-17-009 are included in both panels for comparison. Both panels display two thresholds for $\Lambda$ below which the EFT is invalid, one with $\Lambda = m_{\Phi}$ and the other with the unitarity bound $\Lambda_{95}$ computed in \hyperref[a2]{Appendix B}.}
    \label{exclusions13TeV}
\end{figure*}

With our implementation validated, we proceed to interpret the null results of CMS-EXO-17-009 within our sextet framework. For this task we augment the set of signal samples discussed in \hyperref[s3.1]{Section III.A} with additional samples for sextet masses up to $m_{\Phi} = 5.5\,\text{TeV}$. Recall from \hyperref[s3]{Section III} that these samples include events with muons at both the production and decay stages; these events are discarded due to the muon veto. The results are displayed in \hyperref[recastResultsFig]{Figure 12}, along with the $\sqrt{s}=13\,\text{TeV}$ cross sections first plotted in \hyperref[xsec1]{Figure 1}. The bottom panel of this figure shows the efficiencies of our samples under each final selection (effectively, each signal region) of CMS-EXO-17-009, which we reiterate are tailored to the range of leptoquark masses in \eqref{CMSbins}. All sextet samples are more efficient under cuts tailored for lighter $m_{\text{LQ}}$. We mark the lowest and highest $m_{\text{LQ}}$ signal regions in the legend. The bold curve traces the efficiency of each sextet sample in the signal region that ultimately produces the most stringent upper limit (expected) at 95\% CL on the sextet production cross section. These, along with the observed limits, are displayed in the top panel of \hyperref[recastResultsFig]{Figure 12}. In absolute terms, CMS-EXO-17-009 is less sensitive to our sextet signals than the leptoquark pair production for which it is designed: the imposed limits differ by an order of magnitude. The efficiency loss arises not just from the muon veto but also principally from the $m_{ej}^{\text{min}}$ cuts, since in our production process the two $ej$ pairs chosen following the CMS procedure are not as tightly clustered around the sextet invariant mass as their analogs are around $m_{\text{LQ}}$. Nevertheless, the analysis probes well into the TeV scale both in the sextet mass $m_{\Phi}$ and in the EFT cutoff $\Lambda$, as shown by the selection of three signal cross sections displayed in \hyperref[recastResultsFig]{Figure 12}. For instance, this search disfavors all sextet masses in an effective field theory with a $2\,\text{TeV}$ cutoff (for simplicity, as in \hyperref[s2]{Section II}, we truncate this cross section at $m_{\Phi} = \Lambda$ to preserve EFT validity --- see \hyperref[exclusions13TeV]{Figure 13} below for the partial-wave bound).

%% file: TeX/5_Projections.tex
\section{Projections for LHC Run 3 and beyond}
\label{s5}

Following our discussion in the previous two sections, we now estimate exclusion limits and the discovery potential for our signal at the LHC. We include in this analysis, for direct comparison, the first-generation leptoquark pair search discussed in \hyperref[s4]{Section IV}. The analysis is performed on event samples corresponding to the end of LHC Run 2 (integrated luminosity $\mathcal{L} = 139\,\text{fb}^{-1}$ at center-of-mass energy $\sqrt{s} = 13\,\text{TeV}$) and the planned complete HL-LHC dataset ($\mathcal{L} = 3\,\text{ab}^{-1}$ at $\sqrt{s} = 14\,\text{TeV}$). The cross sections used to normalize the samples at each center-of-mass energy are available in \hyperref[xsec1]{Figure 1} and \hyperref[backgroundTable]{Table I}; the detailed cut-by-cut efficiencies are available for the $\sqrt{s}=13\,\text{TeV}$ samples in Tables \hyperref[cutFlowTable1]{III} and \hyperref[cutFlowTable2]{IV}. For our proposed search, we estimate the discovery threshold --- and, alternatively, the expected limit at 95\% CL if no signal is observed --- in terms of the Asimov approximation for the median signal significance \cite{Cowan:2010js},
\begin{align}\label{significance}
Z_{\text{A}} = \left\lbrace 2\left[(s+b)\ln\,\left(1+\frac{s}{b}\right) - s\right]\right\rbrace^{1/2},
\end{align}
with $s$ and $b$ the signal and background yields after selection, which reduces to the well known form
\begin{align}
Z_{\text{A}} \to \frac{s}{\sqrt{b}} + \mathcal{O}((s/b)^2)
\end{align}
if $s \ll b$, a condition that does not hold for all of our parameter space. For simplicity we neglect signal and background uncertainties; this approximation results in the most optimistic significances. We require the typical values $Z_{\text{A}} = 2$ (5) for exclusion (discovery). Meanwhile, for the leptoquark search CMS-EXO-17-009, we extrapolate the original $\mathcal{L} = 35.9\,\text{fb}^{-1}$ limits to the Run 2 and HL-LHC datasets using the higher-luminosity estimation capabilities of \textsc{MadAnalysis\,5} \cite{Araz_2020}. In this framework, by default, the projected expected limits at 95\% CL are estimated by (i) assuming that signal and background efficiencies are unaffected by any luminosity increase, (ii) rescaling the expected yields according to $\mathcal{L}_{\text{new}}/\mathcal{L}_{\text{original}}$, and (iii) rescaling the systematic (statistical) background uncertainties, which were made available by CMS, according to $\mathcal{L}_{\text{new}}/\mathcal{L}_{\text{original}}$ ($\sqrt{\mathcal{L}_{\text{new}}/\mathcal{L}_{\text{original}}}$). All of these estimates at both luminosities are displayed in \hyperref[exclusions13TeV]{Figure 13}.

Both panels of this figure map the parameter space $(m_{\Phi},\Lambda)$ of our effective field theory with the sextet-quark-lepton couplings \eqref{benchmark} used in the rest of this work. As discussed in \hyperref[s2]{Section II} and \hyperref[s3]{Section III}, this space must be cut off below some minimum $\Lambda$ to preserve the self-consistency of the EFT. We display in \hyperref[exclusions13TeV]{Figure 13} both the naive $\Lambda = m_{\Phi}$ threshold introduced in \hyperref[s2]{Section II} and a lower bound $\Lambda_{95}$ computed in \hyperref[a2]{Appendix B} and given explicitly by \eqref{generalBound}. As alluded to elsewhere in the body of this work, the latter bound is derived from a partial-wave decomposition of the amplitude of the sextet scalar production process(es) and effectively bounds (from above) the center-of-mass energy of such processes for a given cutoff $\Lambda$. These two measures of EFT validity are visibly inequivalent, but they are consistent with each other within an order of magnitude. As we discuss below, most of the parameter space we can probe easily satisfies both bounds.

The left panel of \hyperref[exclusions13TeV]{Figure 13} shows the sensitivity of our search before the final selection is imposed on the invariant mass $m_{\ell_2 j_1j_2}$ of the sextet scalar decay products; \emph{i.e.}, without any attempt to bin the sextet mass. Even at this stage, we see that our search is able to leverage the hard lepton that recoils off of $\Phi$ to improve upon the sensitivity of the CMS leptoquark pair search for TeV-scale sextets and lighter. For heavier sextets, the comparison is more complicated and likely to be qualitatively affected by different handling of signal and background uncertainties. According to our estimates, while our search achieves $\mathcal{O}(100)\,\text{GeV}$ gains in exclusion at each projected luminosity, the extrapolated limits from CMS-EXO-17-009 are strong enough to disfavor the $Z_{\text{A}} = 5$ discovery thresholds for heavy sextets in our search. Massive sensitivity gains are evident in the right panel of \hyperref[exclusions13TeV]{Figure 13}, which shows the results after the final invariant mass cuts. Recall from \hyperref[s3.2]{Section III.B} that the binning is chosen to maximize the signal significance for each simulated signal. Except for sextets lighter than about $500\,\text{GeV}$, for which the best final selections do not meaningfully improve upon the common selections, the $m_{\ell_2 j_1j_2}$ binning is projected to be highly effective, especially for multi-TeV sextets after the full scheduled run of the HL-LHC. This allows the expected sensitivity of our full search to exceed CMS-EXO-17-009 in all directly comparable parameter space. In extreme scenarios, we project HL-LHC exclusions for sextets as heavy as $m_{\Phi} \approx 5.5\,\text{TeV}$ or for very high EFT cutoffs to the tune of $\Lambda \approx 25\,\text{TeV}$. In any case, these projections suggest that the search we have developed would be quite powerful and that the parameter space of our effective theory is ripe for exploration.

\begin{figure}
    \centering
    \includegraphics[scale=0.6]{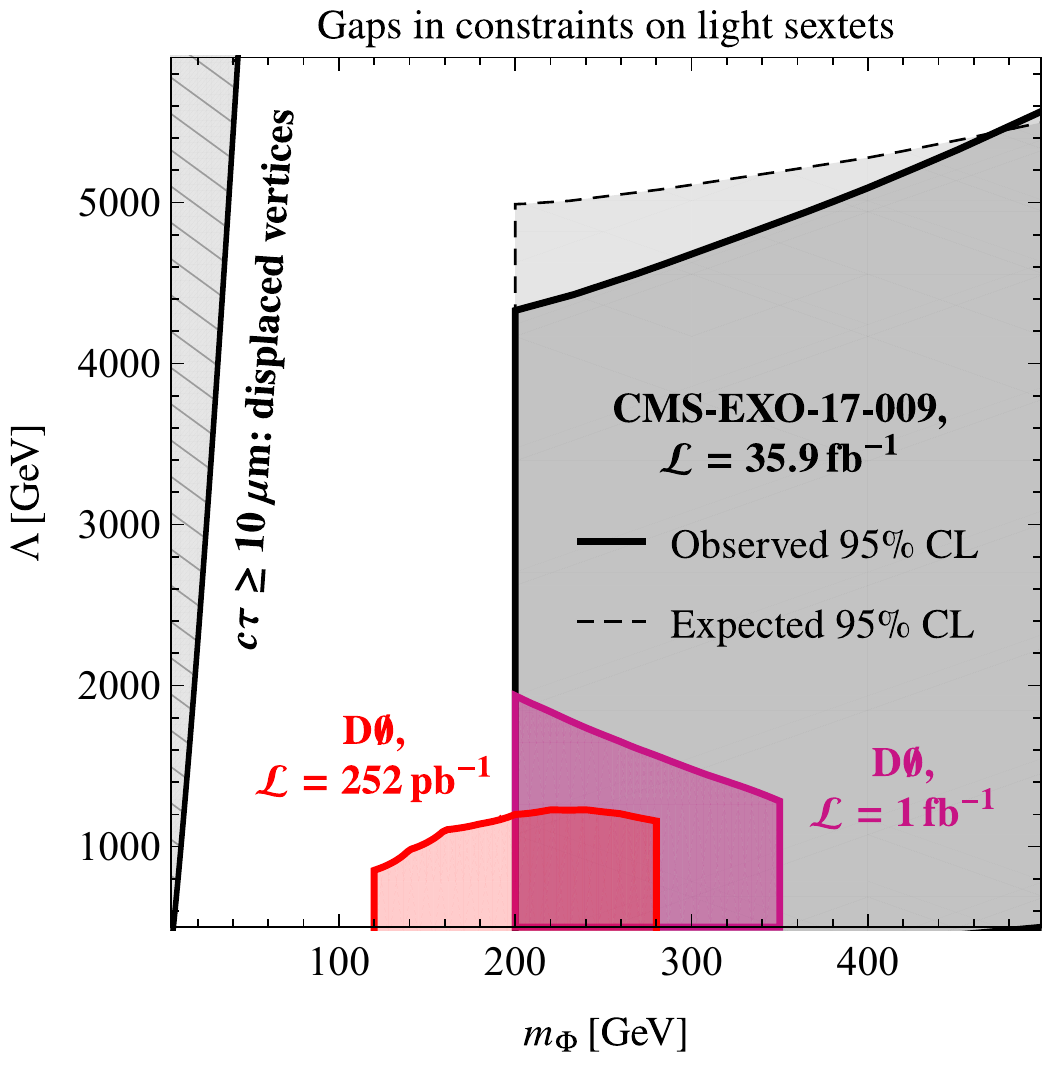}
    \caption{Comparison between LHC and Tevatron ($p\bar{p}$, $\sqrt{s}=1.96\,\text{TeV}$) limits on pair-produced first-generation leptoquarks reinterpreted within our color-sextet scalar EFT. Also displayed is the order-of-magnitude estimate from \hyperref[f1]{Figure 2} of the shortest traveling distance possibly exceeding the resolution of the LHC detectors.}
    \label{oldTevatronLimits}
\end{figure}

Before we conclude, we return to the discussion begun in \hyperref[s2]{Section II} regarding constraints on the very lightest sextet scalars. Despite their high production rates, it appears that a sliver of our EFT parameter space requires a targeted long-lived particle search, and the large light-jet backgrounds at hadron colliders make it difficult to rule out even promptly decaying sextets lighter than about $m_{\Phi} = 200\,\text{GeV}$. The best limits we can find are compared in \hyperref[oldTevatronLimits]{Figure 14}. In addition to CMS-EXO-17-009, which we have discussed at length, we find two searches for first-generation scalar leptoquark pairs ($jj\,e^+e^-$ final states) conducted at the Fermilab Tevatron by the D$\emptyset$ Collaboration. The newer search uses $\mathcal{L} \approx 1\,\text{fb}^{-1}$ of $p\bar{p}$ collisions at $\sqrt{s} = 1.96\,\text{TeV}$ and implements selection criteria fairly similar to those used more recently by CMS \cite{D0:2009jkq}. D$\emptyset$ provides results for leptoquarks as light as $m_{\text{LQ}} \approx 200\,\text{GeV}$, virtually identical to the low end of the $m_{\text{LQ}}$ range targeted by the Run 2 CMS search, and excludes $m_{\text{LQ}} \leq 299\,\text{GeV}$. The older of the two analyses uses a comparatively small $\mathcal{L} = 252\,\text{pb}^{-1}$ dataset \cite{PhysRevD.71.071104}. This search probes leptoquarks as light as $m_{\text{LQ}} = 120\,\text{GeV}$, a bit lighter than the $200\,\text{GeV}$ floor of the other two searches. In the absence of a signal, D$\emptyset$ excludes $m_{\text{LQ}} \leq 241\,\text{GeV}$. To obtain a very rough estimate of the resulting constraints on our sextet signals, we first compute the relevant $\sqrt{s}=1.96\,\text{TeV}$ Tevatron cross sections in \textsc{MG5\texttt{\textunderscore}aMC} version 3.3.1. Then, since (\emph{viz}. Figures \hyperref[LQrecastFig]{11} and \hyperref[recastResultsFig]{12}) our study of CMS-EXO-17-009 shows that \emph{light} sextets and leptoquark pairs have similar limits, we apply the D$\emptyset$ observed limits on the LQ-pair fiducial cross sections directly to our sextet samples, which we expect very slightly overestimates the true limits. Unsurprisingly, we find that the parameter space excluded by D$\emptyset$ in our rough reinterpretation is dwarfed by the region ruled out by the CMS search. More importantly, we find that a fairly large gap exists between these prompt-decay searches and the region where we expect our sextet scalar to be long-lived. This gap may be quite difficult to close at the LHC through standard dijet-resonance searches; alternative approaches in the mold of recent searches for light resonances --- which specifically target collimated decay products characteristic of highly boosted unstable particles \cite{ATLAS:2018hbc} --- may prove more effective.

%% file: TeX/6_Conclusion.tex
\section{Conclusions}
\label{s6}

In this work we have discussed how to search for exotic color-sextet scalars coupling to leptons at the Large Hadron Collider. These scalars can be produced at mass dimension six in association with a hard lepton and subsequently decay to a somewhat softer lepton and two hard jets. After investigating the relevant transverse momentum ($p_{\text{T}}$) and invariant mass distributions of this signal, we have proposed a new search strategy to target its unique lepton kinematics. Our selection criteria include quite stringent cuts on the leading lepton and jet $p_{\text{T}}$. We have assessed the sensitivity of this search to our signal at the LHC; at the same time, we have compared our projections to extrapolations of a CMS search for pair-produced first-generation leptoquarks, which also produce $jj\, \ell^+ \ell^-$ final states, albeit with different momenta. Before we impose final cuts on the invariant mass $m_{\ell_2 j_1j_2}$ of the sextet scalar decay products, we find that the searches are closely competitive for heavier sextet scalars, though our targeting of the hard lepton provides a significant advantage for potential sextets with mass $m_{\Phi} = 1.0\,\text{TeV}$ and lighter. Said final cuts, once imposed, dramatically improve the sensitivity of our search: following the full selection strategy, we expect the $3\,\text{ab}^{-1}$ run of the HL-LHC to exclude sextets as heavy as $5.5\,\text{TeV}$, or EFT cutoffs as high as $25\,\text{TeV}$, \emph{modulo} the order of magnitude of the sextet-quark-lepton couplings. Finally, we have briefly discussed the small parameter space in which our sextet scalar may be long lived, and how a small part of the sizable gap between this region and the range targeted by the analyses discussed in this work might be covered by older searches.

Clearly, there is ample potential for discovery of a semi-leptonically decaying color-sextet scalar at LHC in a broad swath of parameter space. If a signal were observed by our proposed search, it should be possible to discriminate between new-physics hypotheses with different sextet masses through further analysis of $m_{\ell_2 j_1j_2}$. The discovery of a sextet scalar recoiling off of a lepton would constitute very interesting new physics and would immediately motivate ultraviolet completions to the effective operator \eqref{sSmodel}. While we leave a more detailed discussion of the high-energy origin of this model to future work, we note the obvious possibility that heavy leptoquarks could generate the effective operator at loop order. This is just one of many exciting possibilities ready to be explored during the next decade of the LHC program.

%% file: TeX/A_MG5.tex
\section{Simulating the signal in \textsc{MadGraph5\texttt{\textunderscore}aMC@NLO}}
\label{a1}



Here we report some details of the signal event simulation in \textsc{MadGraph5\texttt{\textunderscore}aMC@NLO} 
(\textsc{MG5\texttt{\textunderscore}aMC}) version 3.3.1 \cite{MG5,MG5_EW_NLO}. The present limitations of the color algebra module of \textsc{MG5\texttt{\textunderscore}aMC} make it impossible to simulate $\Phi$ production and decay at the matrix-element level, and \textsc{Pythia\,8} is likewise unable to handle our exotic color structure to perform the decays itself. To circumvent these issues, we have developed a model that produces the same signal as displayed in \hyperref[sig1]{Figure 2} despite relying on a more conventional color structure that can be handled by the computer tools. In this alternative model, the novel scalar (denoted by $\varphi$ to avoid confusion) is a color triplet with weak hypercharge $Y=-1/3$ that interacts with Standard Model fields according to
\begin{align}\label{fake}
\mathcal{L} \supset \frac{1}{\Lambda^2}\,\lambda_{u\ell}^{IX}\,\varphi^{\dagger i}[\bt{t}_{\boldsymbol{3}}^a]_i^{\ \,j}\,(\,\overbar{u^{\text{c}}_{\text{R}}}_{Ii}\,\sigma^{\mu\nu}\ell_{\text{R}X})\, G_{\mu\nu\,a} + \text{H.c.}, 
\end{align}
with $\bt{t}_{\boldsymbol{3}}$ the generators of the fundamental representation of $\mathrm{SU}(3)$. This model permits the signal represented in \hyperref[sig1]{Figure 2} with $\Phi^{\dagger} \to \varphi$ and the reversal of the arrow tracking the flow of analyticity through the scalar propagator. Our strategy is therefore to analyze samples produced using this alternative model but normalized to the cross sections of the sextet model, which differ only by a group-theoretical factor of
\begin{align}
    \frac{\sigma(pp \to \ell^+\Phi^{\dagger})}{\sigma(pp \to \ell^+ \varphi)} = \frac{\tr \bt{J}^{\,s}\, \bar{\!\bt{J}}_{s}}{\tr \bt{t}_{\boldsymbol{3}}^a \bt{t}_{\boldsymbol{3}}^a} = \frac{3}{2}.
\end{align}
The numerator follows from the normalization of the $\boldsymbol{3} \otimes \boldsymbol{6} \otimes \boldsymbol{8}$ Clebsch-Gordan coefficients introduced in \cite{Carpenter:2021rkl}.

%% file: TeX/B_Unitarity.tex
\section{Leading-order perturbative unitarity bound on the effective operator cutoff $\Lambda$}
\label{a2}

The operator \eqref{sSmodel}, being non-renormalizable, can only accurately describe physical processes taking place at energy scales lower than the scale of the degrees of freedom integrated out of some ultraviolet theory in order to produce the infrared model. In the body of this work, we offer two estimates of the range of validity of the effective operator: one simply supposes that the cutoff scale should be identified with the scale of the ultraviolet degrees of freedom, so that $m_{\Phi} \leq \Lambda$; the other is the perturbative unitarity bound \cite{PhysRevD.16.1519,Cohen:2021gdw} that follows from the optical theorem. The latter bound is the focus of this appendix.

In particular, it is well known that the amplitudes $\mathcal{M}_{i\to f}$ of transitions between two-body states $i$ and $f$ satisfies
\begin{align}\label{optical}
2 \Im \mathcal{M}_{i\to i} = \frac{1}{16\pi} \sum_f \beta(s;m_{f_1},m_{f_2}) \int \d \cos \theta\,|\mathcal{M}_{i \to f}|^2
\end{align}
in a unitary theory \cite{PhysRevD.7.3111}. Here $\theta$ is the scattering angle in the center-of-mass reference frame and
\begin{align}
    \beta(s;m_{f_1},m_{f_2}) = \frac{1}{s}\sqrt{[s-(m_{f_1}+m_{f_2})^2][s-(m_{f_1}-m_{f_2})^2]}
\end{align}
is a kinematic function of the final-state particle masses $m_{f_1},m_{f_2}$ and the center-of-mass energy $E_{\text{CM}}^2 = s$ of the process. The perturbative unitarity bound can be derived from \eqref{optical} for a given process \emph{via} a partial-wave decomposition of the corresponding amplitude; namely,
\begin{align}\label{partialWave}
\mathcal{M}_{i\to f} = 8\pi \sum_{J=0}^{\infty} (2J+1)\, \mathcal{T}_{i\to f}^J d_{\mu_i \mu_f}^J(\theta),
\end{align}
with $J$ the angular momentum of each partial wave and $d^J_{\mu_i \mu_f}(\theta)$ the Wigner $d$-functions \cite{edmonds,JACOB2000774} for initial- and final-state total helicities $\mu_i = \lambda_{i_1}-\lambda_{i_2},\ \mu_f=\lambda_{f_1}-\lambda_{f_2}$. In terms of this decomposition, the optical theorem implies for each partial wave
\begin{align}\label{bound1}
2 \Im \mathcal{T}_{i \to i}^J = \beta(s;m_{i_1},m_{i_2})|\mathcal{T}_{i \to i}^J|^2 + \sum_{f \neq i} \beta(s;m_{f_1},m_{f_2})|\mathcal{T}_{i \to f}^J|^2
\end{align}
and, for the inelastic sub-processes \cite{Endo_2014},
\begin{align}\label{bound2}
\sum_{f \neq i} \beta(s;m_{i_1},m_{i_2})\beta(s;m_{f_1},m_{f_2})\,|\mathcal{T}_{i \to f}^J|^2 \leq 1.
\end{align}
We can translate the perturbative unitarity bound \eqref{bound2} into an upper bound on the cutoff $\Lambda$ of the effective operator $\eqref{sSmodel}$ since it generates inelastic $2\to 2$ processes.

We again consider the processes $u_Ig \to \ell_X^+ \Phi^{\dagger}$, whose parton-level cross sections are given by \eqref{xsecAnalytic}. As in \hyperref[s2]{Section II}, we work in the massless-quarks limit. It is clear on inspection of the final state(s) that the production of the sextet scalar $\Phi$ has $J=1/2$. We verify this by computing the transition amplitude $\mathcal{M}(u_Ig \to \ell^+_X \Phi^{\dagger})$ in the center-of-mass frame and with four-component helicity-eigenstate Dirac spinors written according to the Jacob-Wick convention as \cite{JACOB2000774}
\begin{align}\label{explicitSpinors}
u_{\lambda}(p) = \begin{pmatrix}
\sqrt{E+m}\,\chi_{\lambda}\\
2\lambda\sqrt{E-m}\,\chi_{\lambda}\end{pmatrix}\ \ \ \text{and}\ \ \ v_{\lambda}(p)= \begin{pmatrix}
\sqrt{E-m}\,\chi_{-\lambda}\\
-2\lambda\sqrt{E+m}\,\chi_{-\lambda}\end{pmatrix}
\end{align}
with
\begin{align}
    \chi_{\lambda} = \begin{pmatrix}
    \cos \tfrac{\theta}{2}\\
    \e^{\ii \phi} \sin \tfrac{\theta}{2}
    \end{pmatrix}\ \ \ \text{and}\ \ \ \chi_{-\lambda}=\begin{pmatrix}
    -\e^{\ii \phi} \sin \tfrac{\theta}{2}\\
    \cos \tfrac{\theta}{2}\end{pmatrix}
\end{align}
the two-component spinors describing motion in the $\tv{p}$ direction parametrized by polar and azimuthal angles $(\theta,\phi)$. The sextet scalar is produced only for $\lambda_{u_I} = \pm 1/2,\,\lambda_g = \pm 1$ such that $\mu_i = \mu_f = 1/2$. We find that the definite-helicity amplitude in each case takes the form
\begin{align}
    \mathcal{M}(u_I g \to \ell_X^+ \Phi^{\dagger}) = \frac{\lambda_{u\ell}^{IX}}{\Lambda^2}\, \bt{J}^{\,s\,ia}\, f(\hat{s};m_{\Phi},m_{\ell_X}) \cos \frac{\theta}{2},
\end{align}
where
\begin{align}
    f(\hat{s};m_{\Phi},m_{\ell_X}) = -4\hat{s}^{3/4}\left[\sqrt{E_{\ell_X}+m_{\ell_X}} + \sqrt{E_{\ell_X} - m_{\ell_X}}\right]
\end{align}
with
\begin{align}
    E_{\ell_X} = \frac{1}{2\sqrt{\hat{s}}}\,[\hat{s} + m_{\ell_X}^2 - m_{\Phi}^2].
\end{align}
In these expressions $\hat{s}$, the partonic center-of-mass energy, is better described as the (squared) invariant mass $m_{\ell_X^+ \Phi^{\dagger}}$ of the $\ell^+_X \Phi^{\dagger}$ system. The angular factor $\cos \tfrac{\theta}{2}= d_{1/2,1/2}^{1/2}(\theta)$ matches our expectation that sextet scalar production is a pure $J=1/2$ process in the effective model \eqref{sSmodel}. By comparison with \eqref{partialWave}, the perturbative unitarity bound \eqref{bound2} for $u_I g \to \ell_X^+ \Phi^{\dagger}$ can therefore be written as
\begin{align}\label{generalBound}
    1 \geq \frac{3}{(4\pi)^2}\,[\beta(\hat{s};m_{\Phi},m_{\ell_X})]^2\left(\frac{\lambda_{u\ell}^{IX}}{\Lambda^2}\right)^2 |f(\hat{s};m_{\Phi},m_{\ell_X})|^2.
\end{align}
This expression can be inverted to obtain the general lower bound on the EFT cutoff $\Lambda$. In the massless-lepton limit, the perturbative unitarity bound takes the simple form
\begin{align}\label{masslessLeptonBound}
\Lambda \geq \left[6\left(\frac{\lambda_{u\ell}^{IX}}{2\pi}\right)^2\right]^{1/4}(\hat{s} - m_{\Phi}^2)^{1/2}.
\end{align}

In order to obtain a quantitative lower bound on the EFT cutoff $\Lambda$ in our model, we closely follow the procedure adopted in previous works \cite{Endo_2014,Yamamoto:2014pfa} and (conservatively) require that at least 95\% of our sextet signal events satisfy the bound \eqref{generalBound} for any given cutoff $\Lambda$. We specifically use \textsc{MadAnalysis}\,5 to compute the invariant mass of the sextet scalar and the lepton against which it recoils --- using the same signal samples as for the main analysis as described in Sections \hyperref[s3.1]{III.A} and \hyperref[s4]{IV} --- and find the smallest $\Lambda$ for which the efficiency of a cut on $m_{\ell_X^+ \Phi^{\dagger}}$ consistent with \eqref{generalBound} is $95\%$ or greater. The bound as a function of $m_{\Phi}$ is denoted by $\Lambda_{95}$ and is displayed in both panels of \hyperref[exclusions13TeV]{Figure 13}. As mentioned in \hyperref[s3.2]{Section III.B}, the partial-wave unitarity bound $\Lambda_{95}$ motivates our choice of $\Lambda = 3.0\,\text{TeV}$ since a cutoff of that size does not appear to violate unitarity for any of the signals explored in Figures \hyperref[metFig]{6}--\hyperref[reconFig]{9}.